# Supermoiré Chern mosaic in helical trilayer $WSe_2$

Zhenyu Wang[1,2*], Mingjie Zhang[1,2*], Hai Meng[3*], Xiuzhen Li[1,2], Subi Du[1,2], Yaotian Liu[1,2], Siyu Fan[1,2], Xiaofan Shi[1,2], Kenji Watanabe[4], Takashi Taniguchi[5], Wei Yang[1,2], Guangyu Zhang[1,2,6], Bingbing Tong[1,2,7], Guangtong Liu[1,2,7], Li Lu[1,2,7], Jie Shen[1,2], Gang Li[1,2], Jing Song[1], Enke Liu[1], Song Liu[8], Fengcheng Wu[3†], Yang Xu[1,2†]

[1]Beijing National Laboratory for Condensed Matter Physics, Institute of Physics, Chinese Academy of Sciences, Beijing 100190, China
[2]School of Physical Sciences, University of Chinese Academy of Sciences, Beijing 100049, China
[3]School of Physics and Technology, Wuhan University, Wuhan 430072, China
[4]Research Center for Functional Materials, National Institute for Materials Science, 1-1 Namiki, Tsukuba 305-0044, Japan
[5]International Center for Materials Nanoarchitectonics, National Institute for Materials Science, 1-1 Namiki, Tsukuba 305-0044, Japan
[6]Songshan Lake Materials Laboratory, Dongguan 523808, China
[7]Hefei National Laboratory, Hefei 230088, China
[8]Institute of Microelectronics, Chinese Academy of Sciences, Beijing 100029, China

*These authors contributed equally
[†]Email: wufcheng@whu.edu.cn, yang.xu@iphy.ac.cn

**Helically twisted multilayers offer access to moiré physics beyond the single-superlattice paradigm, yet their correlated and topological transport properties remain largely unexplored in semiconductor moiré materials. Here we report magnetotransport measurements of helical trilayer $WSe_2$, in which two coupled moiré patterns relax into a supermoiré landscape composed of inequivalent local topological domains with distinct electronic structures and unequal spatial areas. By electrostatic tuning, we identify a trilayer-hybridized regime where interactions and real-space reconstruction combine to generate a plethora of magnetic and topological states absent in the twisted bilayers. At moiré filling factor $\nu = -1$, we observe a ferromagnetic insulating state that is robust against magnetic field and accompanied by a non-quantized anomalous Hall response ~−4 kΩ. This behaviour is consistent with a time-reversal-symmetry-breaking supermoiré Chern mosaic, in which the Hall response arises from the non-cancelling contributions of local domains with opposite Chern character arranged by the relaxed structure. Under strong magnetic fields, a symmetry-broken Chern insulating state ($C = 1$) emerges near $\nu = -2/3$, displaying a much larger positive Hall response together with strongly enhanced longitudinal resistance, suggestive of field-reconstructed topological minibands and domain-boundary scattering. These results establish relaxed supermoiré semiconductor trilayers as a platform for spatially organized magnetism and topology beyond the bilayer limit.**

**Main:**

Moiré quantum matter has emerged as a powerful platform for engineering electronic phases that are difficult to access in conventional solids, as bandwidths, interaction strengths and band topology can be tuned directly through superlattice geometry[1–5]. Within this context, twisted transition-metal dichalcogenides (TMDCs) have enabled the realization of correlated insulating states[6–9], topological and fractional Chern insulating states[10–15], and unconventional superconductivity[16–19]. However, most experimentally established TMDC moiré phases remain limited to twisted bilayers, where a single moiré potential governs the low-energy electronic structure[1,2]. A central open question is whether introducing multiple moiré length scales can fundamentally reorganize electronic band topology in real space, rather than simply perturbing a miniband. Recent studies of helical trilayer graphene (HTG) have shown that multilayer moiré systems can host electronic structures and phases beyond the bilayer limit, including supermoiré-modulated topological flat bands and correlated states[20–22]. Extending this paradigm to TMDCs opens a qualitatively distinct regime, owing to finite effective mass, reduced lattice symmetry, spin-valley locking[23–26] and electrically tunable layer hybridization[27,28]. Related theories further predict that helical TMDC trilayers can host higher-Chern bands, localized boundary states, and highly tunable supermoiré electronic structures[29,30]. More broadly, when topology becomes spatially dependent, a relaxed supermoiré landscape may support a real-space topological mosaic, in which domains with distinct local topological character are separated by boundary-mode networks[31–34]. If time-reversal symmetry is further broken and the local domains carry different Chern numbers, this texture can generate a Chern mosaic with a net or non-cancelling Hall response[20,21,29,30,35]. Such topological networks share similarities with disordered quantum Hall systems, where insulating longitudinal transport can coexist with a robust and even nearly quantized Hall response, a hallmark of the quantum Hall insulator[36–39]. Supermoiré mosaics therefore provide a natural setting for observing analogous transport behavior, with percolating boundary channels, domain-wall scattering and mesoscopic current partitioning encoded in their real-space topological networks.

Here we approach this problem using helical trilayer $WSe_2$ ($HTWSe_2$) with equal consecutive twisted angles $\theta_{12} = \theta_{23} = \theta$. In this system, moiré patterns from neighboring layer pairs are coupled, producing a relaxed supermoiré landscape composed of unequally sized local domains with distinct electronic structures and valley Chern characters. Using dual-gated magnetotransport measurements, we identify a trilayer-hybridized regime in which supermoiré reconstruction, electronic correlations and topology intertwine to produce magnetic and topological states absent in twisted bilayers. Most prominently, we observe a time-reversal-symmetry-breaking supermoiré Chern mosaic at $\nu = -1$, where interaction-driven ferromagnetism acts on a real-space array of inequivalent Chern domains and gives rise to a non-quantized anomalous Hall response. We further identify a topological mosaic that may host helical boundary-mode networks at $\nu = -2$, as well as a distinct field-induced global $C = 1$ insulating topological state emanating from $\nu = -2/3$. The latter exhibits a large Hall response together with insulating longitudinal transport, suggestive of a field-induced symmetry-broken topological state with quantum-Hall-insulator-like behavior in a patterned supermoiré network.

**Lattice relaxation and topology in helical trilayer $WSe_2$**

We begin by describing the real-space structure, band reconstruction and local topology of $HTWSe_2$. Figure 1a-c illustrate the real- and momentum-space structures of $HTWSe_2$. In such systems, the two adjacent layer pairs generate coupled moiré patterns whose interference produces a longer-scale moiré-of-moiré, or supermoiré, texture[20,21,29,30]. For small $\theta$, the system is characterized by two emergent length scales: the primary moiré period $\lambda_{\mathrm{m}} \approx a/\theta$, where $a$ = 0.3317 nm (ref. 40) is the lattice constant of $WSe_2$, and a much longer supermoiré scale $\lambda_{\mathrm{sm}} \approx a/\theta^2$. In our devices, $\theta \approx 3.9°$, corresponding to $\lambda_{\mathrm{m}} \approx$ 4.9 nm and $\lambda_{\mathrm{sm}} \approx$ 70 nm. In helical moiré systems, lattice relaxation typically reorganizes the nominally incommensurate supermoiré structure into large triangular domains with local moiré periodicity of $\lambda_{\mathrm{m}}$, denoted αβ and βα, respectively[20,21,29,30]. Within each of the moiré-periodic domains, hybridization among states derived from the three rotated monolayers reconstructs the low-energy valence bands, where the *K*-valley states ($K_1$, $K_2$, $K_3$) of the three layers are folded onto the high-symmetry points ($\kappa$, $\gamma$, $\kappa'$) of the moiré Brillouin zone (mBZ)[29,30]. In $HTWSe_2$, as schematically illustrated in the left panel of Fig. 1c, the $\kappa$- and $\kappa'$-valley states derived from the top and bottom layers remain degenerate at zero out-of-plane displacement field ($D$). Meanwhile, the moiré potentials from the two outer layers can add together on the middle layer and raise the valence-band energy of the $\gamma$ valley of the mBZ, indicating middle-layer derived band edge at charge neutrality. Hybridization among these states at their band crossings generates saddle points in the band structure, giving rise to van Hove singularities (vHS) in the density of states (DOS). As $D$ increases, the layer degeneracy is progressively lifted, leading to sequential Lifshitz transitions from a middle-layer polarized band edge (first panel of Fig. 1c) to a bilayer polarized band edge (middle panel of Fig. 1c), and finally to an outer-layer polarized band edge (right panel of Fig. 1c). Because of the spin-valley locking, the spin-up branch at $\kappa$ (red curve) and spin-down branch at $\kappa'$ (blue curve) are related by time reversal symmetry.

There are, however, important distinctions from HTG. In HTG, the two relaxed domain types are $C_{2z}$-symmetry related counterparts, constraining the topological bands within a given valley in the two regions to carry opposite local Berry curvature and Chern numbers[20,21]. In $HTWSe_2$, by contrast, monolayer TMDCs already lack $C_{2z}$, so the relaxed αβ and βα domains are not symmetry partners but instead host genuinely distinct local atomic configurations and band structures. As shown in Fig. 1d-f, the αβ domain contains three atomic-scale high-symmetry stackings, MXX, XMX and XXM, whereas the βα domain contains XMM, MXM and MMX[29,30]. The electronic structures of αβ and βα domains are therefore not constrained to be symmetry-related as in HTG. In Fig. 1g, our calculations based on continuum moiré Hamiltonian for $\theta = 3.9°$ $HTWSe_2$ show a well-isolated topmost moiré miniband whose widths is substantially reduced (roughly by a factor of two) relative to twisted bilayer $WSe_2$ at a comparable twist angle, reflecting the stronger moiré effects in the twisted trilayer. The miniband width of αβ and βα domains differ by about 20% while carrying opposite valley Chern numbers of ±1. For twist angles below ~1.7°, the topological contrast between the αβ and βα domains becomes even stronger in the continuum model, with valley Chern numbers of –2 and −1, respectively (Extended Data Fig. 1). Crucially, our molecular dynamics simulation indicates that the αβ domain is energetically more favored over the βα domains, so the lattice relaxation produces unequally sized distorted triangular domains, with the αβ regions substantially enlarged relative to the βα regions, as schematically shown in Fig. 1d and discussed in Methods. This pronounced relaxation effect points to an unexpectedly strong role of supermoiré reconstruction in $HTWSe_2$ even the twist angle is as large as ∼3.9°. The first moiré miniband in the shrunk αα domains (circled by a dashed green circle in Fig. 1d) remains topologically trivial throughout the calculated range of twist angles from 1.5° to 4°. Together, these results support a physical picture in which locally reconstructed domains carry distinct topological characters and are arranged by the relaxed supermoiré texture into a real-space Chern mosaic[20,29,30,35,41–44].

**Layer occupation and emergent ferromagnetism**

Figure 1h and 1i show the transport device schematic and optical micrograph (see more details about device fabrication in Methods and Extended Data Fig. 2). Adjacent layers of $HTWSe_2$ are rotated by the nearly identical angles, $\theta_{12} \approx \theta_{23} \approx 3.9°$. The sample is encapsulated by hexagonal boron nitride (hBN) and dual-gated by graphite top and bottom gates, enabling independent control of the carrier density $n$ (hence moiré filling factor $v$) and displacement field $D$. Here $v$ is defined with respect to the moiré, rather than the supermoiré, lattice. A local contact gate electrostatically dopes the $WSe_2$ at the prepatterned Pt contact regions, yielding Ohmic contacts down to carrier densities as low as $2 \times 10^{11}$ $cm^{-2}$ at low temperatures. The device is etched into a Hall-bar geometry.

Figure 2a and 2b show the longitudinal resistance $R_{xx}$ as a function of filling factor $v$ and displacement field $D$, measured at zero magnetic field and base temperature (nominal $T = 10$ mK), together with the corresponding layer-occupation phase diagram. Six distinct regions are clearly resolved. In the low-$|v|$ sector, three single-layer-polarized regions, $I_1$, $I_2$ and $I_3$, appear in different $|D|$ windows, where the subscript denotes the layer index. Among them, the middle-layer-polarized region $I_2$ occupies a triangular area at relatively small $|D|$ and $|v| < 0.53$, whereas the top- and bottom-layer-polarized regions $I_1$ and $I_3$ emerge at larger $|D|$. Temperature-dependent Shubnikov-de Haas (SdH) oscillations (Extended Data Fig. 3) indicate a slightly larger effective mass in region $I_2$ (~$0.38m_e$, with $m_e$ being the free electron mass) than in regions $I_1$ and $I_3$ (~$0.29m_e$). Together with the angle-dependent SdH measurements (Extended Data Fig. 4), these relatively small effective masses support an origin of these states from the monolayer $K$ and $K$' valleys rather than the $\Gamma$ valley. At intermediate $D$, two bilayer-polarized regions, $II_{12}$ and $II_{23}$, dominate the transport. Their behavior is broadly consistent with that of twisted bilayer $WSe_2$, exhibiting generalized Wigner crystal states at $v = -1/4$ and $-1/3$, as well as an insulating state at $v = -1$ that is later shown to be antiferromagnetic[9,16–19,45]. The sharpness and close alignment of these correlated insulating states in the top-middle ($II_{12}$) and middle-bottom ($II_{23}$) sectors indicate excellent twist-angle matching between $\theta_{12}$ and $\theta_{23}$, with a difference below ~0.03° (Extended Data Fig. 5). Between the two bilayer-polarized regions lies a trilayer-hybridized region (III) at $|v| > 0.53$, in which all three layers contribute to transport. The most prominent feature of region III is a triangular high-resistance region centered near $\nu = -1$ and $D = 0$, where $R_{xx}$ exceeds 50 kΩ at its maximum, along with an additional extended resistance peak at $\nu = -2$ that will be discussed later.

As shown in Fig. 2d-h, interaction-driven ferromagnetism develops throughout this triangular high-resistance region and extends through a fishtail-like feature towards $\nu = -2$ at larger $|D|$. Figure 2d and 2e map the magnetoresistance, defined as $MR = [R_{xx}(B) - R_{xx}(0)]/R_{xx}(0)$, and the Hall resistance $R_{xy}$, respectively, in the $\nu$-$D$ plane at a small perpendicular magnetic field of $B$ = 0.1 T. Representative magnetic-field sweeps at five selected points, marked in Fig. 2d, are shown in Fig. 2g and 2h. As summarized in the schematic magnetic phase diagram in Fig. 2f, three types of ferromagnetic behavior can be identified. Near $\nu = -1$ and $D = 0$, we observe a ferromagnetic insulating state characterized by a strongly enhanced $R_{xx}$ and a large non-quantized anomalous Hall response of ~4 kΩ. Surrounding this phase is a broad region exhibiting a sharp negative magnetoresistance (NMR) peak centered near zero magnetic field, with $R_{xx}$ reduced by nearly 70% for $|B|< 0.1$ T while the anomalous Hall response remains weak. In magnetic systems, such low-field NMR is commonly associated with a multidomain structure near magnetization reversal. Here it is consistent with a supermoiré multidomain regime carrying different magnetization orientations, such that domain-wall scattering is suppressed as a small field aligns the domains. At larger carrier density and displacement field, extending towards $\nu \approx -2$, we identify a more itinerant ferromagnetic metallic region that follows the fishtail-like feature and exhibits small Hall jumps (<~100 Ω) and negligible NMR.

To understand the above findings, in Fig. 2c we compare the experimental phase diagrams with the calculated DOS as a function of filling factor $\nu$ and interlayer potential difference $\Delta$ for the αβ domain at $\theta \approx 3.9°$. The calculation captures the main features of the measured phase diagram, aside from the correlation-driven insulating states that are absent in the single-particle picture. In particular, the vHS in the regions $II_{12}$ and $II_{23}$ crosses $\nu = -1$ at a relatively large $|\Delta|$, corresponding well to the observed $\nu = -1$ correlated insulators in these regions. Furthermore, the vHS evolve continuously into region III, intersect at $\Delta = 0$ near $\nu = -1$, and then disperse towards larger $|\Delta|$ with increasing hole density. This evolution accounts for the fishtail-like high-resistance background of the trilayer-hybridized region and closely tracks the extent of the magnetic phase diagram. The correspondence is especially notable for the itinerant ferromagnetic metallic region at larger $|D|$ and higher hole density, consistent with a DOS-enhanced, possibly Stoner-type ferromagnetic instability[45,46]. By contrast, the ferromagnetic insulating state at $\nu = -1$ and $D = 0$ represents the strongest instability in region III, accompanied by a large anomalous Hall response and suggestive of a possible connection to the supermoiré Chern-mosaic picture[20,21,29,30,35,41–44].

**Supermoiré Chern mosaic**

To further investigate the nature of the ferromagnetic insulating state at $\nu = -1$ in region III, we compare its temperature and magnetic-field dependences in Fig. 3 with those of the insulating states in the bilayer-hybridized regions at finite $D$. Figure 3a shows $R_{xx}$ as a function of displacement field $D$ and temperature $T$ at fixed $\nu = -1$. The insulating state near $D = 0$ persists and is further strengthened upon increasing magnetic field (Fig. 3b), with the activation gap increasing from ~1 meV at $B = 0$ T to ~4 meV at 6 T (inset of Fig. 3b and Extended Data Fig. 6a), indicating that this state is robust against $B$ field. Moreover, the corresponding $R_{xy}(B)$ exhibits a pronounced anomalous Hall response with clear hysteresis and stochastic, Barkhausen-like jumps at low temperature (Fig. 3c), both of which gradually weaken upon warming. The Curie temperature, $T_c$ ~9 K (see more discussion in Methods and Extended Data Fig. 7), marks the onset of the nonlinearity in $R_{xy}(B)$ and coincides with the resistance upturn of $R_{xx}(T)$ at $B = 0$. By contrast, the $\nu = -1$ insulating state at finite displacement field (in region $II_{12}$ or $II_{23}$), exemplified here by $D = 0.29$ V/nm, shows the opposite magnetic-field dependence. As seen in Fig. 3d and 3e, this state is rapidly suppressed by magnetic field and evolves into a metallic state, while the corresponding activation gap decreases markedly with increasing field and vanishes above ~6 T (inset of Fig. 3e and Extended Data Fig. 6b). This behavior is consistent with the antiferromagnetic insulator previously reported in twisted bilayer $WSe_2$ at similar twist angles[9,47,48]. It is also reminiscent of a recent study in twisted bilayer $WSe_2$, where a correlated insulator was progressively weakened by perpendicular magnetic field and, upon the formation of conjugate electron- and hole-Landau levels, evolved into an oscillatory sequence of quantum spin Hall (QSHI) and excitonic insulating (EI) states[47].

A possible microscopic picture for the intriguing ferromagnetic insulating state at $\nu = -1$ and $D = 0$ is illustrated schematically in Fig. 3f. In the relaxed supermoiré structure, the two dominant inequivalent stacking domains, αβ and βα, carry opposite local Chern character for a given polarized valley/spin flavor. Without time-reversal-symmetry breaking, the two time-reversed flavors remain degenerate and their Hall responses cancel. Interaction-driven flavor polarization can lift this degeneracy and generate spontaneous magnetization, thereby exposing the opposite local Chern numbers of the αβ and βα domains, $C = -1$ and $C = +1$, respectively. Because lattice relaxation substantially enlarges the αβ regions relative to the βα regions, the Hall contributions from the two types of domains do not fully cancel and produce a net anomalous Hall response of a few kΩ. At the same time, chiral topological edge states propagate along the boundaries between neighboring topological domains, circulating counter-clockwise around the αβ domains (red arrows) and clockwise around the βα domains (blue arrows), and undergo scattering at the junctions of the shrunken αα domains (light green regions). Within this picture, the ferromagnetic insulating state at $\nu = -1$ and $D = 0$ can be understood as an interaction-polarized supermoiré Chern mosaic, naturally accounting for the coexistence of a large but non-quantized anomalous Hall response and a sizeable, yet non-divergent, longitudinal resistance[21,35,41].

By contrast, the resistance peak near $\nu = -2$ at small $|D|$ shows no evidence of time-reversal-symmetry breaking. This filling corresponds to a paired spin-valley configuration in which the first moiré minibands associated with the αβ and βα domains are filled in both time-reversed flavors. The two domains can then be regarded as QSH-like local regions with gapped bulks, vanishing charge Chern number and finite spin/valley Chern character, whose interfaces may support counterpropagating helical boundary modes. The $\nu = -2$ state therefore represents a time-reversal-preserved topological mosaic of the relaxed supermoiré landscape, analogous to the proposal in TMDC heterobilayers[35,49].

**Field-induced symmetry-broken Chern insulating state (SBCIS)**

We next examine the magnetic-field evolution of transport in $HTWSe_2$ along the $D = 0$ line, which cuts through the center of the trilayer-hybridized region. Figure 4a and 4b show the symmetrized longitudinal resistance $R_{xx}$ and antisymmetrized Hall resistance $R_{xy}$ as functions of magnetic field $B$ and filling $\nu$ at $D = 0$ and $T = 15$ mK. The corresponding Wannier diagram is shown in Fig. 4c. Starting from the band edge at $\nu = 0$, well-developed quantum Hall states emerge with increasing magnetic field, with $R_{xx}$ approaching zero and $R_{xy}$ exhibiting clear quantization, underscoring the high quality of the device. These Landau levels mainly originate from the middle-layer-derived states in region $I_2$. At higher hole densities, however, additional field-induced features appear within the trilayer-hybridized regime III. In particular, extra Landau-level-like or Hofstadter-like branches develop near $\nu = -1$, the insulating state at $\nu = -1$ persists to high fields, and the resistance peak at $\nu = -2$ disappears above $B \sim 12$ T.

Most notably, a topological feature with Středa index $C = +1$, extrapolating to $\nu = -2/3$ at $B = 0$ T, emerges above $B \approx 4$ T, as indicated by the red line in Fig. 4c. To further probe this state, Fig. 4d and 4e present magnetic-field sweeps at different temperatures and fixed filling factor near $\nu = -2/3$. A field-driven metal-insulator transition occurs around $|B| \approx 4$ T. At higher fields, both $R_{xx}$ and $|R_{xy}|$ increase upon cooling, in marked contrast to the behavior expected for a clean, uniform quantum Hall state or Chern insulator. At lowest temperatures, $R_{xx}$ exceeds 100 kΩ, while $|R_{xy}|$ becomes larger than 18 kΩ, approaching the quantized value $h/e^2 \sim 25.8$ kΩ. This phenomenology is distinct from previously reported symmetry-broken Chern insulators, where $R_{xx}$ typically develops a minimum or is strongly suppressed inside the topological gap[5,50–52].

Although both states show insulating behavior accompanied by large Hall responses, the $(C, \nu) = (+1, -2/3)$ state is unlikely to be a simple continuation of the zero-field $\nu = -1$ Chern mosaic. This distinction is most directly reflected in their Hall polarity and magnitude: the $\nu = -1$ ferromagnetic state shows a negative anomalous Hall response of only a few kΩ, whereas the high-field state follows a $C = +1$ Středa trajectory and displays a much larger positive Hall response, with $|R_{xy}| > 18$ kΩ. Within the Chern-mosaic picture, the negative anomalous Hall polarity at $\nu = -1$ arises from the non-cancelling contributions of the two inequivalent αβ and βα domains, which carry opposite local Chern character (−1 and +1, respectively) for the interaction-driven spin-valley polarized state. The high-field $C = +1$ state therefore points to a distinct field-induced topological reconstruction. In this scenario, magnetic field reshapes the trilayer minibands and brings vHS close to the fractional filling $v = -2/3$ near $D = 0$ (Extended Data Fig. 8), thereby enhancing interactions that may favor moiré-translation-symmetry breaking, possibly in the form of charge density wave (CDW) order, as illustrated schematically in Fig. 4f. Band folding by the enlarged unit cell can open up a gap at this fractional filling, redistributing Berry curvature among the resulting subbands[53,54] to stable a global $C = +1$ response. Within this picture, different regions of the relaxed supermoiré landscape may undergo distinct field-driven topological reconstructions, while the domain boundaries remain weak links that form metallic or weakly localized scattering networks[35]. The resulting state accounts for the coexistence of a large Hall response and a strongly enhanced longitudinal resistance, reminiscent of the quantum Hall insulator in disordered two-dimensional electron systems, where the Hall conductance remains large or even quantized while dissipative transport is governed by an inhomogeneous percolative network[35–37,39]. We therefore refer to this state as a field-induced symmetry-broken Chern insulating state (SBCIS) in the trilayer-hybridized regime.

## Conclusion

In summary, our measurements reveal $HTWSe_2$ as a multilayer moiré system where relaxed supermoiré reconstruction gives rise to correlated and topological transport phenomena beyond the bilayer paradigm. These findings establish helical TMDC trilayers as a platform for organizing topology and magnetism in real space through hierarchical supermoiré reconstruction. Future studies that vary the two consecutive twist angles ($\theta_{12}$, $\theta_{23}$) may uncover new supermoiré textures and topological phases, including possible routes towards correlated topological superconductivity when bilayer-hybridized superconducting regimes are brought into proximity with trilayer-hybridized magnetic or Chern-mosaic regimes in the $n$–$D$ phase space. Such supermoiré engineering may further enable more robust Chern insulators, fractional Chern mosaics and fractionalized excitations[20,35,42–44].

## Methods

### Device Fabrication

Devices were fabricated using a standard dry-transfer technique[55]. Atomically thin flakes of hexagonal boron nitride (hBN), graphite and $WSe_2$ were mechanically exfoliated from bulk crystals onto $SiO_2$/Si substrates. As summarized in Extended Data Fig. 2, fabrication proceeded in three main steps. First, the bottom-gate structure was prepared by sequentially picking up hBN and few-layer graphite with a polycarbonate (PC) stamp and releasing them onto a $SiO_2$/Si substrate. Pt contact electrodes (6 nm) were then defined on this structure by electron-beam lithography followed by electron-beam deposition. After metal deposition, the Pt contact regions were cleaned by contact-mode atomic force microscopy (AFM) to remove polymer residues. Next, a large monolayer $WSe_2$ flake was mechanically cut into three pieces using an AFM tip. With a separate PC stamp, hBN, few-layer graphite, hBN and the three $WSe_2$ pieces were then sequentially picked up. During the pickup of successive $WSe_2$ layers, the stage was rotated by a prescribed angle relative to the previous layer, thereby forming a helically stacked twisted trilayer $WSe_2$ structure. The assembled heterostructure was subsequently released onto the substrate with pre-patterned Pt electrodes. Finally, contact-gate electrodes (3/30 nm Cr/Au) were defined by electron-beam lithography and deposited by electron-beam evaporation. The device was then patterned into a standard Hall-bar geometry by reactive-ion etching.

### Transport measurements

Electrical transport measurements were carried out in three cryostats: two dilution refrigerators and one cryostat equipped with a $^3$He insert. One dilution refrigerator was an Oxford TLM system equipped with an 18 T superconducting magnet and a nominal base temperature of 15 mK. Silver-powder filters and RC filters were installed at low temperature to suppress high-frequency noise. The second dilution refrigerator was an Oxford Triton system equipped with a 9-1-1 T vector superconducting magnet and a nominal base temperature of 10 mK. This system used room-temperature π-filters together with low-temperature RC filters. Measurements at intermediate temperatures were performed in a cryostat equipped with a $^3$He insert (down to 280 mK), a 16 T superconducting magnet, and low-temperature RC filters. All transport measurements were performed using standard low-frequency lock-in techniques with excitation frequencies below 20 Hz, using NF LI-5650 or NF LI-5640 lock-in amplifiers. To minimize electron-heating effects, the excitation current was kept below 10 nA throughout the measurements.

### Twist-angle determination

In the dual-gate configuration, the carrier density $n$ and displacement field $D$ are independently controlled by the top- and bottom-gate voltages, $V_{tg}$ and $V_{bg}$, according to

$$n = \frac{C_{tg}V_{tg} + C_{bg}V_{bg}}{e} - n_{offset}$$

$$D = \frac{C_{bg}V_{bg} - C_{tg}V_{tg}}{2\varepsilon_0} - D_{offset}$$

where $e$ is the elementary charge, $\varepsilon_0$ is the vacuum permittivity, and $C_{tg}$ and $C_{bg}$ are the top- and bottom-gate capacitances per unit area, respectively. The gate capacitances were calibrated from the Landau fan diagrams at high magnetic fields. In our device, pronounced features at moiré fillings $\nu = -1/4, -1/3, -1$ and $-2$ in the dual-gate maps provide reliable markers for determining the moiré carrier density $n_M$. Shubnikov-de Haas (SdH) oscillations at high magnetic fields were used as an independent consistency check for the gate-capacitance calibration and the extracted twist angle. Furthermore, linecuts in the $\nu$-$D$ plane at $D = 0.29$ V/nm and $-0.29$ V/nm was used to estimate the twist angles of the top-middle and middle-bottom layer interfaces. Taking the $R_{xx}$ peaks at filling $\nu = -1$ in the bilayer-polarized regions as the reference for the moiré densities yield the extracted twist angles are 3.92° and 3.95°, respectively, indicating excellent twist-angle uniformity across the device (Extended Data Fig. 5).

**Angle-dependence and effective mass**

To identify the valley character of the carriers in the low $|D|$ and low $|\nu|$, we measured the angle dependence of the SdH oscillations by rotating the sample with respect to the magnetic field at $D = 0$ V/nm and $n = -0.91 \times 10^{12}$ cm$^{-2}$, corresponding to the middle-layer-polarized region $I_2$ (Extended Data Fig. 4). Here the tilt angle $\theta$ is defined as the angle between the magnetic field and the sample normal, such that the out-of-plane field component is $B_\perp = B\cos\theta$. When the longitudinal resistance is plotted as a function of $B_\perp$, the curves measured at different tilt angles nearly collapse onto the same traces. This shows that the magnetoresistance and SdH are governed predominantly by the perpendicular magnetic field, whereas the in-plane component has little effect within the accessible range. This weak in-plane-field dependence, together with the location of the measurement point in region $I_2$, indicates that the relevant carriers mainly originate from the $K$-valley states of the middle layer, with no contribution from $\Gamma$-valley states. This assignment is consistent with the modelling used in the main text, which focuses on the middle-layer-derived $K$-valley states.

We further extracted the effective mass $m^*$ from the temperature dependence of the SdH oscillation amplitude using the standard Lifshitz-Kosevich formula (Extended Data Fig. 3),

$$\Delta R \propto \frac{X}{\sinh X},$$

$$X = \frac{2\pi^2 k_B T m^*}{\hbar e B}$$

Here, $\Delta R$ is the SdH oscillation amplitude, $k_B$ is the Boltzmann constant, $T$ is the temperature, $m^*$ is the cyclotron effective mass, $\hbar$ is the reduced Planck constant, $e$ is the elementary charge, and $B$ is the magnetic field at which the oscillation amplitude is evaluated. We find that the effective mass depends strongly on the layer character of the carriers. At relatively large $|D|$, where the carriers mainly reside in the top or bottom layer, the extracted effective mass is close to the monolayer value after spin polarization. By contrast, at $D = 0$, where the carriers are dominated by the middle layer, the effective mass is enhanced, indicating stronger band renormalization in the middle-layer-dominated states.

**Thermal activation analysis**

The activation gaps shown in Fig. 3b and 3e were extracted from the temperature dependence of $R_{xx}$ measured at $\nu = -1$ and fixed displacement field. For an insulating state governed by thermal activation across an energy gap $\varepsilon$, the resistance follows

$$R_{xx}(T) = R_0 \exp\left(\frac{\varepsilon}{2k_{\mathrm{B}}T}\right)$$

where $R_0$ is a prefactor and $k_{\mathrm{B}}$ is the Boltzmann constant. We analyzed the data empirically using Arrhenius plots of $\ln R_{xx}$ as a function of $1/T$. Linear fits were performed over the temperature interval displaying activated behaviors, as highlighted by the dashed lines in Extended Data Fig. 6, and the activation gap $\varepsilon$ was determined from the fitted slope. At low temperatures, data points showing saturation or strong departure from linearity were excluded from the fit.

**Extraction of Curie temperature**

The Curie temperature $T_{\mathrm{C}}$ is extracted using an Arrott-plot-like analysis[21,56,57]. Specifically, we plot $R_{xy}^2$ as a function of $|B/R_{xy}|$ at different temperatures (Extended Data Fig. 7), taking as an approximate measure of the magnetization. In this representation, a linear extrapolation of the high-field regime yields an intercept whose sign characterizes the magnetic state: a positive (negative) intercept corresponds to a ferromagnetic (paramagnetic) phase. The intercept approaches zero at $T \approx 9$ K, from which we estimate a Curie temperature of $T_{\mathrm{C}} \approx 9$ K.

**Numerical calculations**

The relaxed structure of HTWSe$_2$ in Fig. 1c is obtained by using a machine learned force field (MLFF) trained for twisted trilayer WSe$_2$ combined with molecular dynamics (MD). We use the DeepMD-kit package[58,59] to train the MLFF with DFT generated dataset and the LAMMPS software[60,61] to perform MD simulation for a commensurate twist angle $\theta_{12} = \theta_{23} \approx 3.89°$. We will report a more detailed description of the MLFF in a separate work.

We use a continuum model for the calculation of the band structures and DOS of αβ, βα and αα HTWSe$_2$ at valley $K$. The continuum Hamiltonian is given by:

$$H_\phi(\boldsymbol{r}) = \begin{pmatrix} h^{(3)}(\boldsymbol{k}) + \Delta^{32}_{\phi,+}(\boldsymbol{r}) - \Delta & T^{32}_\phi(\boldsymbol{r}) & \\ \left[T^{32}_\phi(\boldsymbol{r})\right]^* & h^{(2)}(\boldsymbol{k}) + \Delta^{32}_{\phi,-}(\boldsymbol{r}) + \Delta^{21}_{\phi,+}(\boldsymbol{r}) & T^{21}_\phi(\boldsymbol{r}) \\ & \left[T^{21}_\phi(\boldsymbol{r})\right]^* & h^{(1)}(\boldsymbol{k}) + \Delta^{21}_{\phi,-}(\boldsymbol{r}) + \Delta \end{pmatrix},$$

$$h^{(3)}(\boldsymbol{k}) = -\frac{\hbar^2(\boldsymbol{k}-\boldsymbol{\kappa}')^2}{2m^*};\ h^{(2)}(\boldsymbol{k}) = -\frac{\hbar^2\boldsymbol{k}^2}{2m^*};\ h^{(1)}(\boldsymbol{k}) = -\frac{\hbar^2(\boldsymbol{k}-\boldsymbol{\kappa})^2}{2m^*},$$

$$\Delta^{32}_{\phi,\pm}(\boldsymbol{r}) = 2V \sum_{j=1,3,5} \cos(\boldsymbol{g}_j \cdot \boldsymbol{r} \pm \psi + \phi),$$

$$\Delta^{21}_{\phi,\pm}(\boldsymbol{r}) = 2V \sum_{j=1,3,5} \cos(\boldsymbol{g}_j \cdot \boldsymbol{r} \pm \psi - \phi),$$

$$T^{32}_\phi(\boldsymbol{r}) = w[1 + \exp(-i\boldsymbol{g}_2 \cdot \boldsymbol{r} + i\phi) + \exp(-i\boldsymbol{g}_3 \cdot \boldsymbol{r} - i\phi)],$$

$$T_{\phi}^{21}(\boldsymbol{r}) = w[1 + \exp(-i\boldsymbol{g}_2 \cdot \boldsymbol{r} - i\phi) + \exp(-i\boldsymbol{g}_3 \cdot \boldsymbol{r} + i\phi)],$$

where $\phi = \frac{2}{3}\pi$, $-\frac{2}{3}\pi$ and 0 correspond to αβ, βα and αα stacking, respectively. Here $\boldsymbol{\kappa} = \frac{4\pi}{3\lambda_{\mathrm{m}}}(0,1)$ and $\boldsymbol{\kappa}' = \frac{4\pi}{3\lambda_{\mathrm{m}}}(0,-1)$, where $\lambda_{\mathrm{m}} \approx a/\theta$ is the moiré period and $a$ =0.3317 nm (ref. 40) is the lattice constant of monolayer layer $WSe_2$. $\boldsymbol{g}_j$ are moiré reciprocal lattice vectors given by $\boldsymbol{g}_j \equiv \sqrt{3}|\boldsymbol{\kappa}|(\cos[(j-1)\pi/3], \sin[(j-1)\pi/3])$. $\Delta$ is the interlayer potential difference. The continuum model parameters are taken as[29] $\psi = 77.02°, w = 14.32$ meV, $V = 5.966$ meV, and $m^* = 0.43\, m_e$.

**References**

1. Mak, K. F. & Shan, J. Semiconductor moiré materials. *Nat. Nanotechnol.* **17**, 686–695 (2022).
2. Li, B., Qiu, W.-X., Wu, F. & MacDonald, A. H. Quantum phases in twisted homobilayer transition metal dichalcogenides. *Natl. Sci. Rev.* **13**, nwaf570 (2026).
3. Andrei, E. Y. & MacDonald, A. H. Graphene bilayers with a twist. *Nat. Mater.* **19**, 1265–1275 (2020).
4. Balents, L., Dean, C. R., Efetov, D. K. & Young, A. F. Superconductivity and strong correlations in moiré flat bands. *Nat. Phys.* **16**, 725–733 (2020).
5. Nuckolls, K. P. & Yazdani, A. A microscopic perspective on moiré materials. *Nat. Rev. Mater.* **9**, 460–480 (2024).
6. Regan, E. C. *et al.* Mott and generalized Wigner crystal states in $WSe_2/WS_2$ moiré superlattices. *Nature* **579**, 359–363 (2020).
7. Tang, Y. *et al.* Simulation of Hubbard model physics in $WSe_2/WS_2$ moiré superlattices. *Nature* **579**, 353–358 (2020).
8. Xu, Y. *et al.* Correlated insulating states at fractional fillings of moiré superlattices. *Nature* **587**, 214–218 (2020).
9. Wang, L. *et al.* Correlated electronic phases in twisted bilayer transition metal dichalcogenides. *Nat. Mater.* **19**, 861–866 (2020).
10. Li, T. *et al.* Quantum anomalous Hall effect from intertwined moiré bands. *Nature* **600**, 641–646 (2021).
11. Cai, J. *et al.* Signatures of fractional quantum anomalous Hall states in twisted $MoTe_2$. *Nature* **622**, 63–68 (2023).
12. Zeng, Y. *et al.* Thermodynamic evidence of fractional Chern insulator in moiré $MoTe_2$. *Nature* **622**, 69–73 (2023).
13. Park, H. *et al.* Observation of fractionally quantized anomalous Hall effect. *Nature* **622**, 74–79 (2023).
14. Xu, F. *et al.* Observation of integer and fractional quantum anomalous Hall effects in twisted bilayer $MoTe_2$. *Phys. Rev. X* **13**, 031037 (2023).
15. Foutty, B. A. *et al.* Mapping twist-tuned multiband topology in bilayer $WSe_2$. *Science* **384**, 343–347 (2024).
16. Xia, Y. *et al.* Superconductivity in twisted bilayer $WSe_2$. *Nature* **637**, 833–838 (2025).
17. Guo, Y. *et al.* Superconductivity in 5.0° twisted bilayer $WSe_2$. *Nature* **637**, 839–845 (2025).
18. Xia, Y. *et al.* Bandwidth-tuned Mott transition and superconductivity in moiré $WSe_2$. *Nature* **650**, 585–591 (2026).
19. Guo, Y. *et al.* Angle evolution of the superconducting phase diagram in twisted bilayer $WSe_2$. *Nature* **652**, 622–627 (2026).
20. Devakul, T. *et al.* Magic-angle helical trilayer graphene. *Sci. Adv.* **9**, eadi6063 (2023).
21. Xia, L.-Q. *et al.* Topological bands and correlated states in helical trilayer graphene. *Nat. Phys.* **21**, 239–244 (2025).

22. Hoke, J. C. *et al.* Imaging supermoiré relaxation in helical trilayer graphene. *Nat. Mater.* (2026).
23. Xiao, D., Liu, G.-B., Feng, W., Xu, X. & Yao, W. Coupled spin and valley physics in monolayers of $MoS_2$ and other group-VI dichalcogenides. *Phys. Rev. Lett.* **108**, 196802 (2012).
24. Xu, X., Yao, W., Xiao, D. & Heinz, T. F. Spin and pseudospins in layered transition metal dichalcogenides. *Nat. Phys.* **10**, 343–350 (2014).
25. Manzeli, S., Ovchinnikov, D., Pasquier, D., Yazyev, O. V. & Kis, A. 2D transition metal dichalcogenides. *Nat. Rev. Mater.* **2**, 17033 (2017).
26. Mak, K. F., Xiao, D. & Shan, J. Light–valley interactions in 2D semiconductors. *Nat. Photonics* **12**, 451–460 (2018).
27. Wu, F., Lovorn, T., Tutuc, E., Martin, I. & MacDonald, A. H. Topological insulators in twisted transition metal dichalcogenide homobilayers. *Phys. Rev. Lett.* **122**, 086402 (2019).
28. Devakul, T., Crépel, V., Zhang, Y. & Fu, L. Magic in twisted transition metal dichalcogenide bilayers. *Nat. Commun.* **12**, 6730 (2021).
29. Nakatsuji, N., Kawakami, T., Tateishi, H., Kato, K. & Koshino, M. Moiré band engineering in twisted trilayer $WSe_2$. *Commun. Mater.* **6**, 274 (2025).
30. Choi, J. D. *et al.* Higher Chern bands in helical homotrilayer transition metal dichalcogenides. *Phys. Rev. B* **112**, 205122 (2025).
31. San-Jose, P. & Prada, E. Helical networks in twisted bilayer graphene under interlayer bias. *Phys. Rev. B* **88**, 121408 (2013).
32. Rickhaus, P. *et al.* Transport through a network of topological channels in twisted bilayer graphene. *Nano Lett.* **18**, 6725–6730 (2018).
33. Huang, S. *et al.* Topologically protected helical states in minimally twisted bilayer graphene. *Phys. Rev. Lett.* **121**, 037702 (2018).
34. Xu, S. G. *et al.* Giant oscillations in a triangular network of one-dimensional states in marginally twisted graphene. *Nat. Commun.* **10**, 4008 (2019).
35. Bhattacharjee, S., May-Mann, J., Kwan, Y. H., Devakul, T. & Sharpe, A. Mesoscopic transport in a Chern mosaic. *arXiv:2604.08654* (2026).
36. Zhang, S.-C., Kivelson, S. & Lee, D.-H. Zero-temperature Hall coefficient of an insulator. *Phys. Rev. Lett.* **69**, 1252–1255 (1992).
37. Shimshoni, E. & Auerbach, A. Quantized Hall insulator: Transverse and longitudinal transport. *Phys. Rev. B* **55**, 9817–9823 (1997).
38. Hilke, M. *et al.* Experimental evidence for a two-dimensional quantized Hall insulator. *Nature* **395**, 675–677 (1998).
39. Shimshoni, E. The quantized Hall insulator: A 'quantum' signature of a 'classical' transport regime? *Mod. Phys. Lett. B* **18**, 923–943 (2004).
40. Mounet, N. *et al.* Two-dimensional materials from high-throughput computational exfoliation of experimentally known compounds. *Nat. Nanotechnol.* **13**, 246–252 (2018).

41. Grover, S. *et al.* Chern mosaic and Berry-curvature magnetism in magic-angle graphene. *Nat. Phys.* **18**, 885–892 (2022).
42. Guerci, D., Mao, Y. & Mora, C. Chern mosaic and ideal flat bands in equal-twist trilayer graphene. *Phys. Rev. Res.* **6**, L022025 (2024).
43. Datta, A., Guerci, D., Goerbig, M. O. & Mora, C. Helical trilayer graphene in magnetic field: Chern mosaic and higher Chern number ideal flat bands. *Phys. Rev. B* **110**, 075417 (2024).
44. Kwan, Y. H., Tan, T. & Devakul, T. Fractional Chern mosaic in supermoiré graphene. *Phys. Rev. Res.* **7**, L032070 (2025).
45. Knüppel, P. *et al.* Correlated states controlled by a tunable van Hove singularity in moiré $WSe_2$ bilayers. *Nat. Commun.* **16**, 1959 (2025).
46. Stoner, E. C. Collective electron ferromagnetism. *Proc. R. Soc. Lond. Ser. A. Math. Phys. Sci.* 165, 372–414 (1938).
47. Han, Z. *et al.* Quantum oscillations between excitonic and quantum spin Hall insulators in moiré $WSe_2$. *arxiv:2509.19287* (2025).
48. Ghiotto, A. *et al.* Stoner instabilities and Ising excitonic states in twisted transition metal dichalcogenides. *arXiv:2405.17316* (2024).
49. Tong, Q. *et al.* Topological mosaics in moiré superlattices of van der Waals heterobilayers. *Nat. Phys.* **13**, 356–362 (2017).
50. Spanton, E. M. *et al.* Observation of fractional Chern insulators in a van der Waals heterostructure. *Science* **360**, 62–66 (2018).
51. Saito, Y. *et al.* Hofstadter subband ferromagnetism and symmetry-broken Chern insulators in twisted bilayer graphene. *Nat. Phys.* **17**, 478–481 (2021).
52. Yu, J. *et al.* Correlated Hofstadter spectrum and flavour phase diagram in magic-angle twisted bilayer graphene. *Nat. Phys.* **18**, 825–831 (2022).
53. Su, R. *et al.* Moiré-driven topological electronic crystals in twisted graphene. *Nature* **637**, 1084–1089 (2025).
54. Xiang, H. *et al.* Continuously tunable anomalous Hall crystals in rhombohedral heptalayer graphene. *arXiv:2502.18031* (2025).
55. Wang, L. *et al.* One-dimensional electrical contact to a two-dimensional material. *Science* **342**, 614–617 (2013).
56. Arrott, A. Criterion for ferromagnetism from observations of magnetic isotherms. *Phys. Rev.* **108**, 1394–1396 (1957).
57. Chiba, D. *et al.* Electrical control of the ferromagnetic phase transition in cobalt at room temperature. *Nat. Mater.* **10**, 853–856 (2011).
58. Wang, H., Zhang, L., Han, J. & E, W. DeePMD-kit: A deep learning package for many-body potential energy representation and molecular dynamics. *Comput. Phys. Commun.* **228**, 178–184 (2018).
59. Zeng, J. *et al.* DeePMD-kit v2: A software package for deep potential models. *J. Chem. Phys.* **159**, 054801 (2023).
60. Thompson, A. P. *et al.* LAMMPS - a flexible simulation tool for particle-based

materials modeling at the atomic, meso, and continuum scales. *Comput. Phys. Commun.* **271**, 108171 (2022).

61. Plimpton, S. Fast Parallel Algorithms for short-range molecular dynamics. *J. Comput. Phys.* **117**, 1–19(1995).

**Author Contributions:**

Y.X. and Z.W. conceived and designed the experiments. Z.W. fabricated the device and performed the transport measurements, with assistance from M.Z, S.D, Y.L and S.F. X.L, W.Y. and G.Z. assisted the metal deposition. X.S, B.I, G.L, L.L., J.S. and E.L. assisted the measurements in ultra-low temperature. Z.W., M.Z. and Y.X. analyzed the data. H.M. and F.W performed theoretical studies. S.L. grew the bulk $WSe_2$ crystals. K.W. and T.T. grew the bulk hBN crystals. Z.W., M.Z., H.M., F.W. and Y.X. co-wrote the manuscript. All authors discussed the results and commented on the manuscript.

**Competing interests:** The authors declare no competing interests.

## Figures

**Fig. 1 | Lattice symmetry, reconstructed supermoiré, and device architecture of helically twisted trilayer $WSe_2$. a,** Schematic of a helical trilayer $WSe_2$ ($HTWSe_2$) structure. The orange, blue, and green spheres represent W atoms in different layers ($l$ = 1, 2, and 3), while the grey spheres represent Se atoms. The twist angles between adjacent layers are $\theta_{12}$ and $\theta_{23}$; In this device, $\theta_{12} = \theta_{23} \approx 3.9°$, forming a chiral stacking configuration. **b,** Showing the three rotated monolayer Brillouin zones whose $K$ valleys ($K_1$, $K_2$, $K_3$) relax onto the same line in the local moiré-periodic domains and fold to the high-symmetry points ($\kappa$, $\gamma$, $\kappa'$) in the moiré Brillouin zone (mBZ). **c,** Schematic evolution of the low-energy moiré bands in $HTWSe_2$ as a function of displacement field $D$, showing the transition from a layer-symmetric band structure to layer-polarized bands. Red and blue branches represent the $K$ (spin-up) and $K'$ (spin-down) valleys, respectively. **d,** Relaxed real-space supermoiré pattern in $HTWSe_2$, consisting primarily of two large inequivalent domains shaded in red and blue. Pink and blue dots denote the MM stacking sites of moiré 12 and moiré 23, formed by top-middle and middle-bottom layers, respectively. The scale bar is 25 nm. **e,f,** Representative schematics of the two inequivalent local domains, βα and αβ, illustrating the local moiré structure and the corresponding atomic stacking configurations. At the local moiré scale, filled and open triangles denote XM stack and MX stack domains, respectively. **g,** Calculated band structures for the βα and αβ domains with $\theta_{12} = \theta_{23} = 3.9°$, the first isolated mini bands show distinct bandwidth and opposite Chern character, whose spatial coexistence on the supermoiré scale gives rise to a Chern-mosaic landscape. **h,** Schematic of the $HTWSe_2$ Hall-bar device. Gate voltages ($V_{tg}$ and $V_{bg}$) are applied to the top and bottom graphite (Gr) gates, enabling independent control of the carrier density $n$ and displacement field $D$. A local Cr/Au contact gate is employed to highly dope the $WSe_2$ near the Pt electrodes and ensure low contact resistances. **i,** Optical micrograph of the $HTWSe_2$ device. Scale bar, 10 μm.

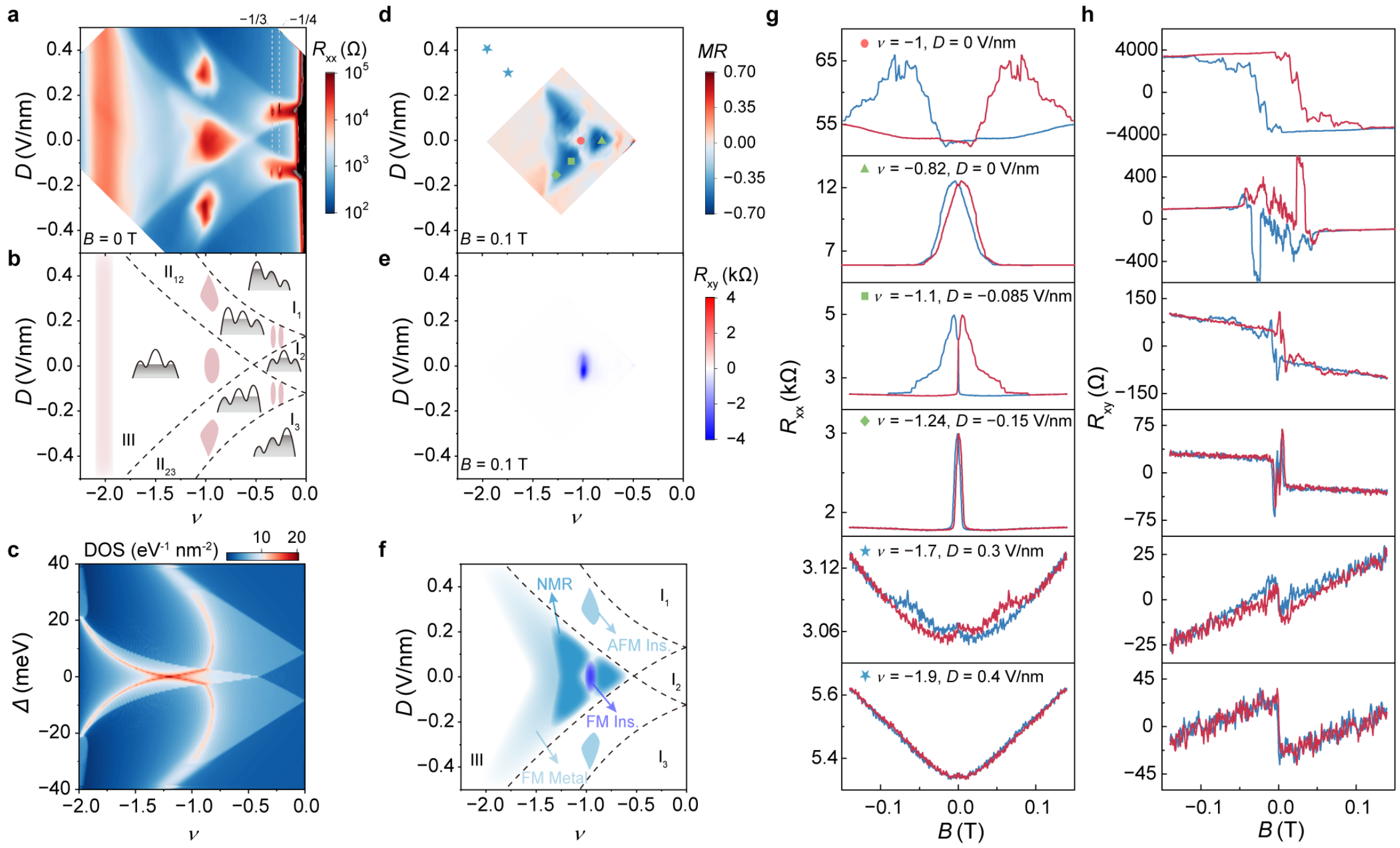

**Fig. 2 | Layer-occupation phase diagram and emergent ferromagnetism in the trilayer-hybridized regime. a,** Longitudinal resistance $R_{xx}$ as a function of filling factor $\nu = n/n_m$ and displacement field $D$, measured at zero magnetic field, where $n_m$ is the moiré superlattice density. **b,** Layer-occupation phase diagram constructed from **a**. Regions I, II, and III correspond predominantly to mono-, bi-, and tri-layer occupation, respectively; subscripts indicate the corresponding layer combination. Insets illustrate representative low-energy band structures and the associated Fermi-level positions for spin-up branch. Pink regions denote insulating states. **c,** Calculated DOS for αβ domain as a function of filling factor $\nu$ and interlayer potential difference $\Delta$, showing clear layer-polarization features. The DOS maxima correspond to van Hove singularities (vHS). **d,e,** Magnetoresistance $MR$ (**d**) and Hall resistance $R_{xy}$ (**e**) mapped in the $\nu$-$D$ plane at $B$ = 0.1 T. The blue regions in **d** show large negative magnetoresistance (NMR). A pronounced Hall signal only develops near $\nu \approx -1$ and $D = 0$. **f,** Schematic magnetic phase diagram inferred from transport, highlighting three magnetic regions in the trilayer-hybridized sector: a ferromagnetic insulator (FM Ins.) near $\nu = -1$, a surrounding negative-magnetoresistance region (NMR), and a more extended ferromagnetic metal (FM metal). Additional antiferromagnetic insulator (AFM Ins.) phases are shown in the bilayer-polarized regions. **g,h,** Magnetic-field dependence of $R_{xx}$ (**g**) and $R_{xy}$ (**h**) at representative points marked in **d**. Red and blue traces denote up- and down-sweeps of the magnetic field, respectively. These points can be divided into three groups, with the pink, green, and blue markers corresponding to the three distinct magnetic regimes summarized in **f**. All data in this figure were measured at nominal $T$ = 10 mK, except for the last two sets of traces in **g** and **h**, which were measured at $T$ = 0.28 K.

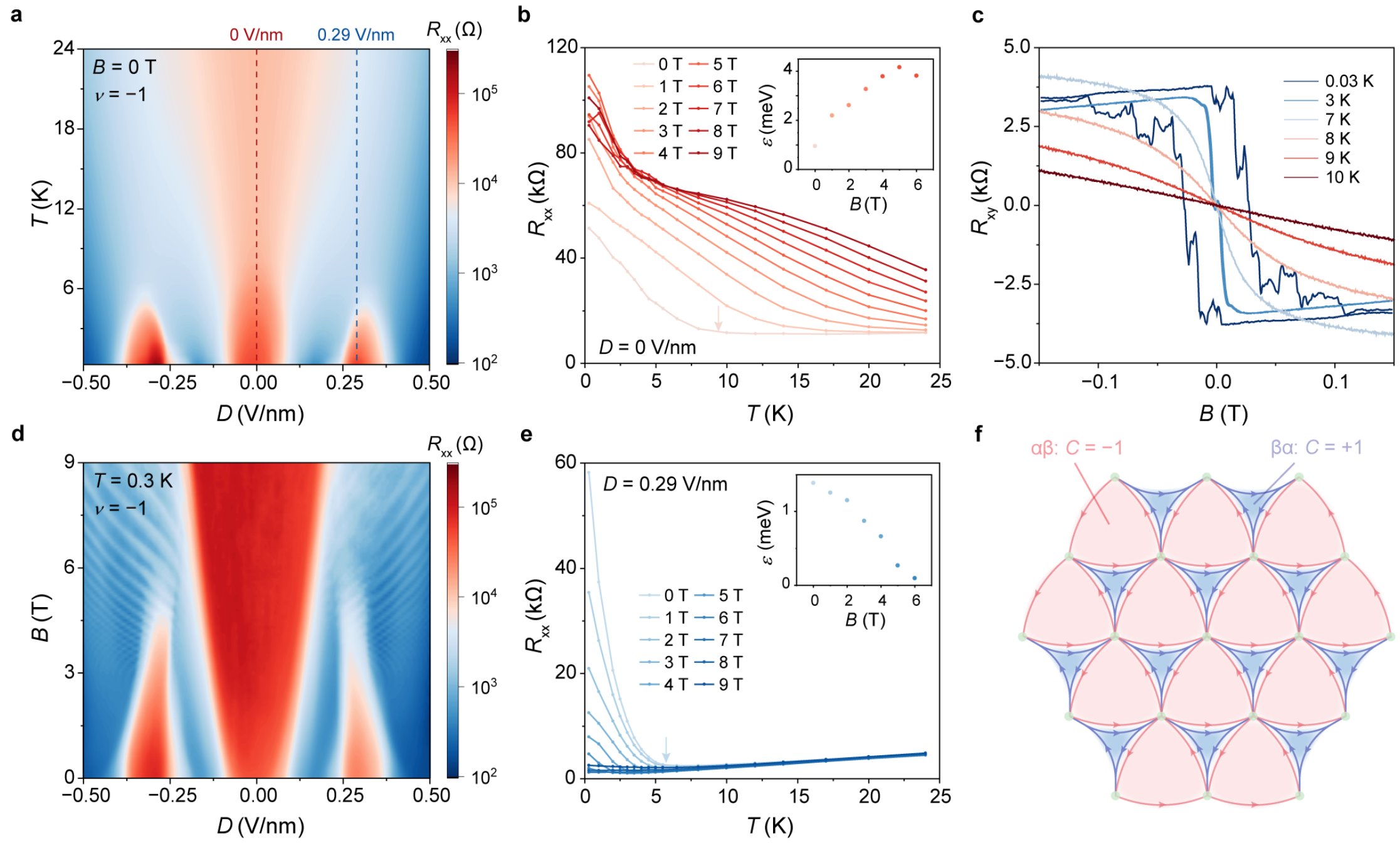

**Fig. 3 | Nature of the $\nu = -1$ insulating states and supermoiré Chern mosaic. a,** Longitudinal resistance $R_{xx}$ as a function of displacement field $D$ and temperature $T$ at $\nu = -1$ and $B = 0$ T. Dashed lines mark the two representative displacement fields $D = 0$ V/nm and 0.29 V/nm, used for the line cuts in **b** and **e**, respectively. **b,** Temperature dependence of $R_{xx}$ at $D = 0$ V/nm under selected magnetic fields. The resistance remains strongly enhanced with increasing field, indicating that the insulating state is robust against magnetic field. Arrow marks the temperature at which $R_{xx}$ begins to turn up upon cooling at $B = 0$ T, corresponding to the onset of the insulating state. Inset, activation gap $\varepsilon$ extracted from thermal activation analysis as a function of magnetic field. **c,** Hall resistance $R_{xy}$ as a function of magnetic field at $D = 0$ V/nm and $\nu = -1$ for selected temperatures. A pronounced anomalous Hall response with hysteresis is observed at low temperatures and gradually weakens upon warming. **d,** Longitudinal resistance $R_{xx}$ as a function of $D$ and $B$ at $\nu = -1$ and $T =$ 0.3 K. **e,** Temperature dependence of $R_{xx}$ at $D = 0.29$ V/nm under selected magnetic fields. In contrast to the state at $D = 0$ V/nm, this insulating state is rapidly suppressed by magnetic field and evolves towards metallic behavior, suggesting a distinct magnetic ground state. Arrow marks the corresponding insulating onset temperature at $B = 0$ T. Inset, magnetic-field dependence of the extracted activation gap $\varepsilon$. **f,** Schematic real-space picture of the relaxed supermoiré Chern mosaic and its chiral edge-state network. Red- and blue-shaded distorted triangles denote the two dominant inequivalent stacking domains, αβ and βα, respectively. Arrows along the domain boundaries indicate the propagation direction of the edge states. The opposite chirality reflects the opposite Chern character of the two domains.

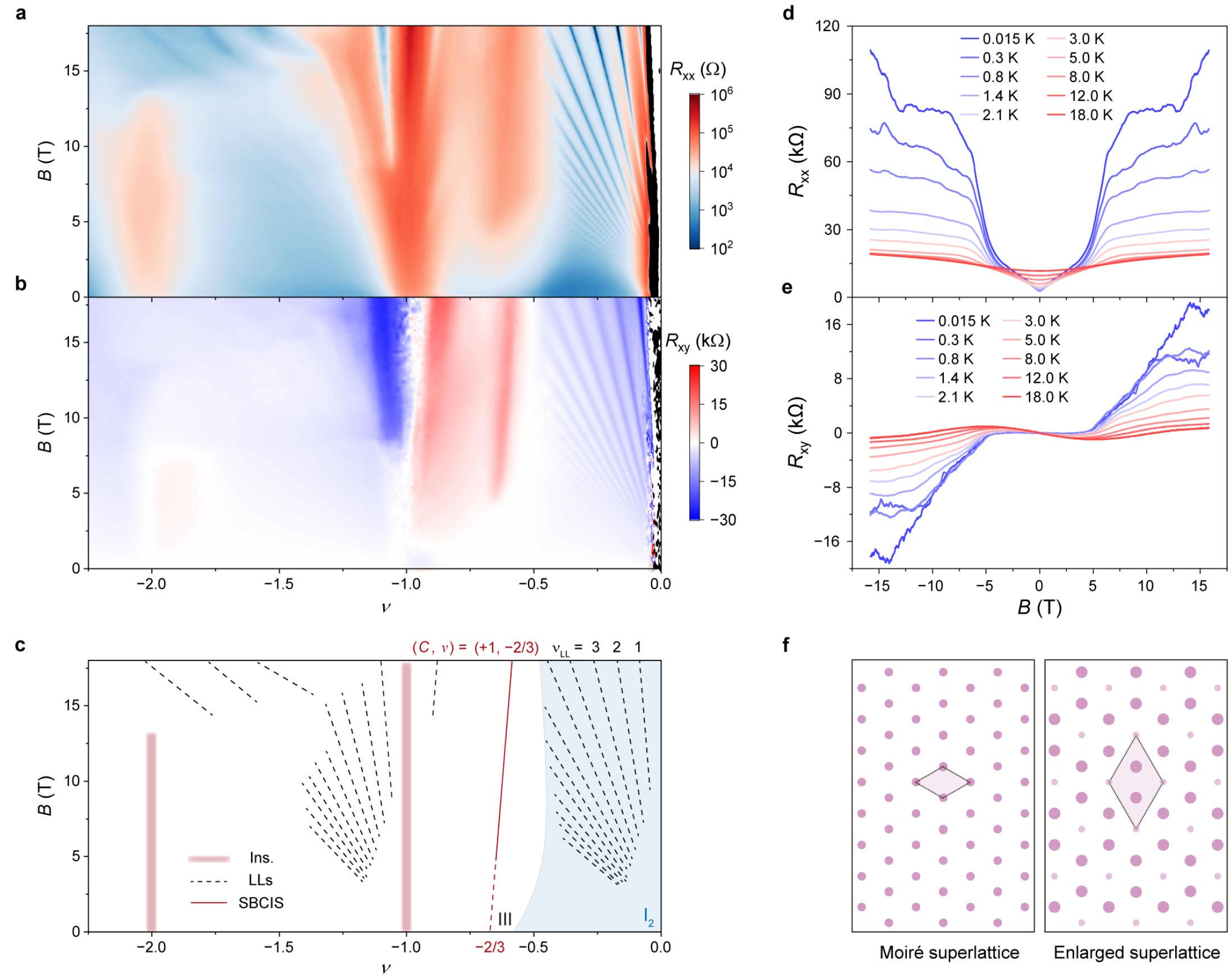

**Fig. 4 | Magnetic-field evolution and field-induced phase transition. a,b,** Symmetrized $R_{xx}$ (**a**) and anti-symmetrized $R_{xy}$ (**b**) as a function of magnetic field $B$ and filling factor $\nu$ at $D = 0$ V/nm, measured at $T = 15$ mK. Distinct field-evolving features emerge near $\nu = -1$ and $\nu = -2/3$. **c,** Wannier diagram summarizing the main field-dependent features in **a** and **b**. The pink shading highlights the insulating state at $\nu = -1$ and $\nu = -2$. The red line denotes the expected Středa slope of the $\nu = -2/3$ state, consistent with $C = +1$. The black dashed lines indicate the Landau levels (LL). **d,e,** Magnetic-field dependence of symmetrized $R_{xx}$ (**d**) and anti-symmetrized $R_{xy}$ (**e**) at selected temperatures, taken at fixed filling $\nu = -0.615$. This cut intersects the $\nu = -2/3$ state. A field-induced metal-insulator transition occurs around $B$ ~4 T. At higher fields, both $R_{xx}$ and $R_{xy}$ increase upon cooling. **f,** Schematic illustration of the real-space lattice reconstruction. Left, the moiré superlattice sites. Right, the formation of a $\sqrt{3} \times \sqrt{3}$ CDW superlattice under $B$ fields.

## Extended Data Figures

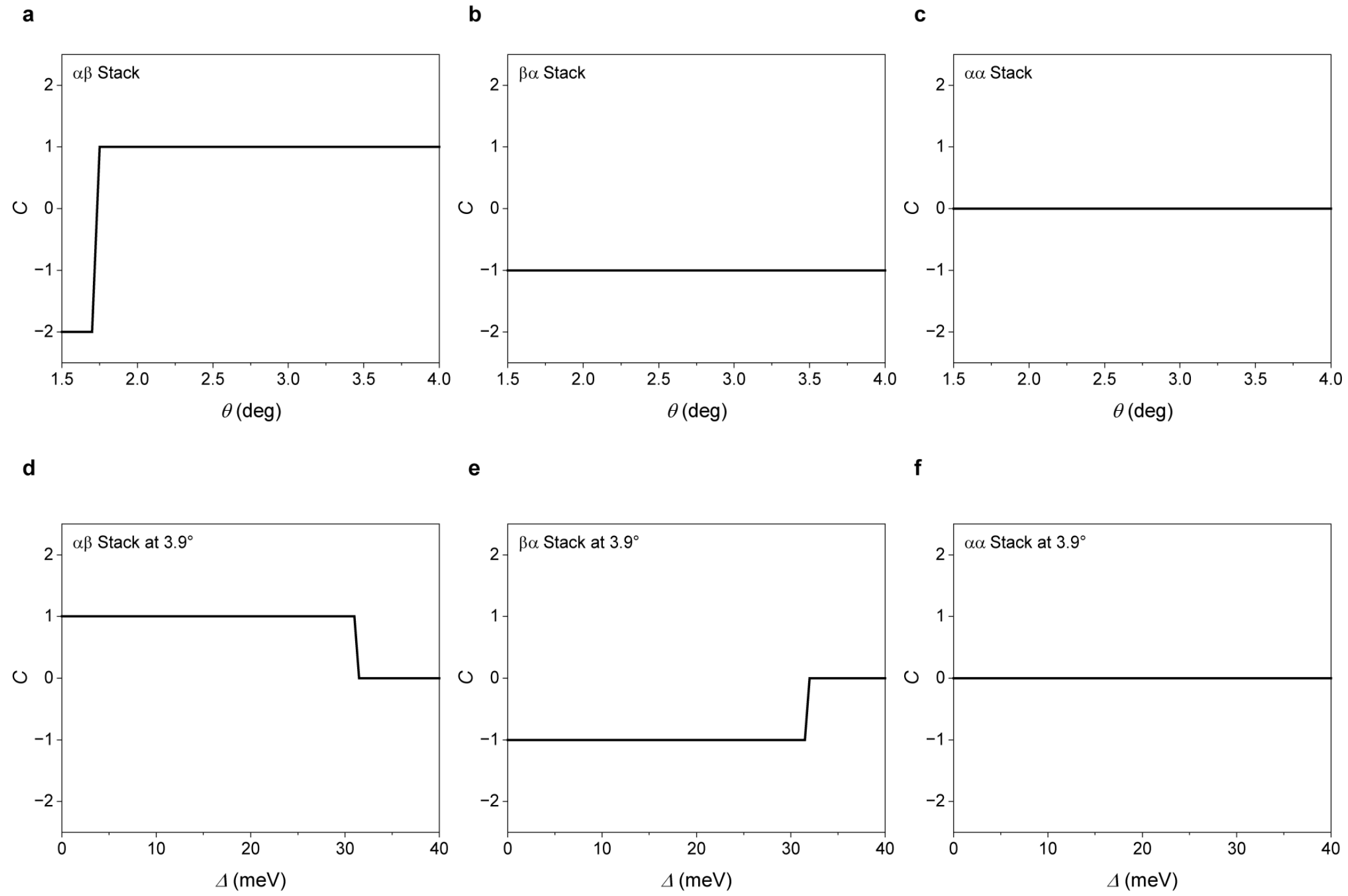

**Extended Data Fig. 1 | Stacking-dependent evolution of Chern number. a-c,** Calculated Chern number $C$ in $K$ valley versus twist angle $\theta$ for the first moiré miniband in the αβ, βα, and αα stacking regions, respectively. The inequivalent αβ and βα stackings are not symmetry constrained, and can therefore exhibit completely different Chern numbers at the same twist angle. **d-f,** Calculated $C$ versus interlayer potential difference $\Delta$ at fixed $\theta$ = 3.9° for the three stacking regions. As $\Delta$ increases, corresponding to increasing displacement field, $C$ is driven to zero, indicating a transition to a topologically trivial band regime at large displacement field.

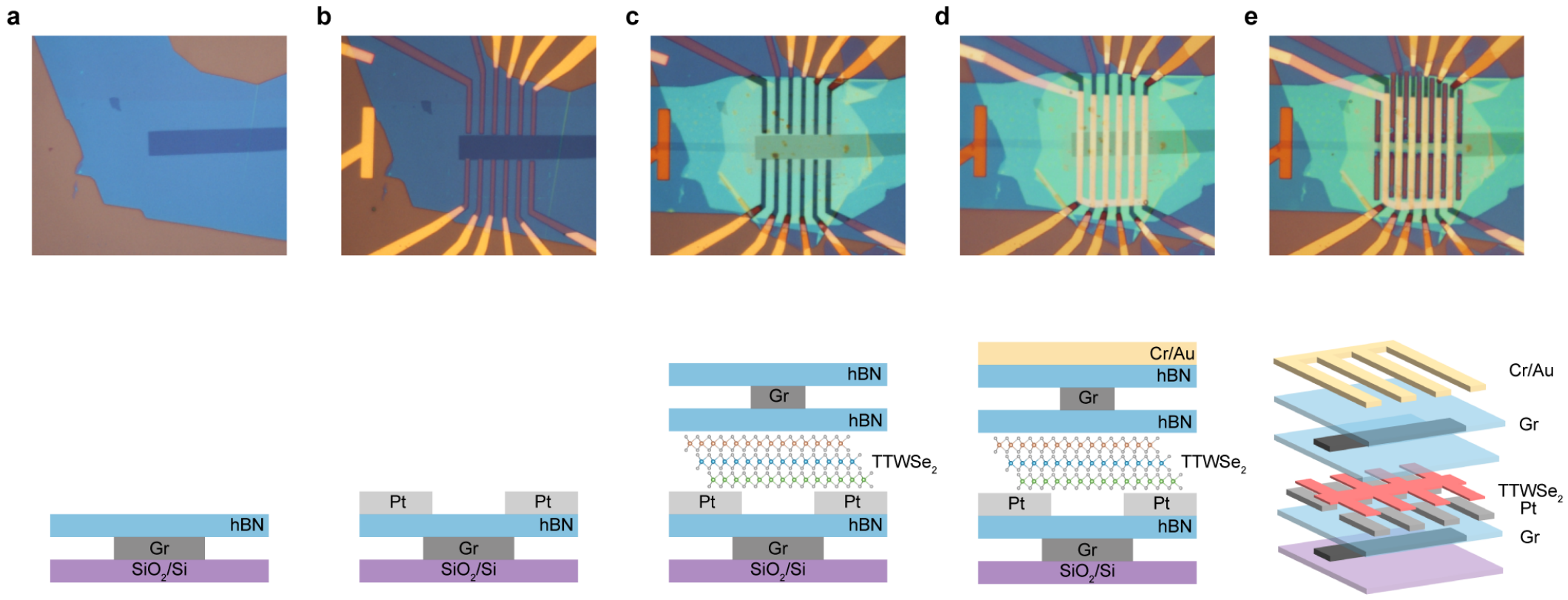

**Extended Data Fig. 2 | Step-by-step fabrication of the $HTWSe_2$ device. a-e,** Top: optical micrographs of representative fabrication steps. Bottom: corresponding schematic illustrations. **a,** Preparation of the bottom-gate substrate, consisting of graphite encapsulated by hBN on $SiO_2$/Si. **b,** Definition of Pt electrodes on the bottom-gate structure. **c,** Transfer of the hBN/graphite/hBN-encapsulated $HTWSe_2$ stack onto the pre-patterned Pt electrodes. **d,** Deposition of Cr/Au contact-gate electrodes. **e,** Schematic of the final device. The heterostructure was subsequently patterned into a Hall-bar geometry by reactive ion etching (RIE). The final device consists of a $HTWSe_2$ channel contacted by Pt electrodes and sandwiched between graphite top and bottom gates, with Cr/Au contact gates used to electrostatically dope the contact regions.

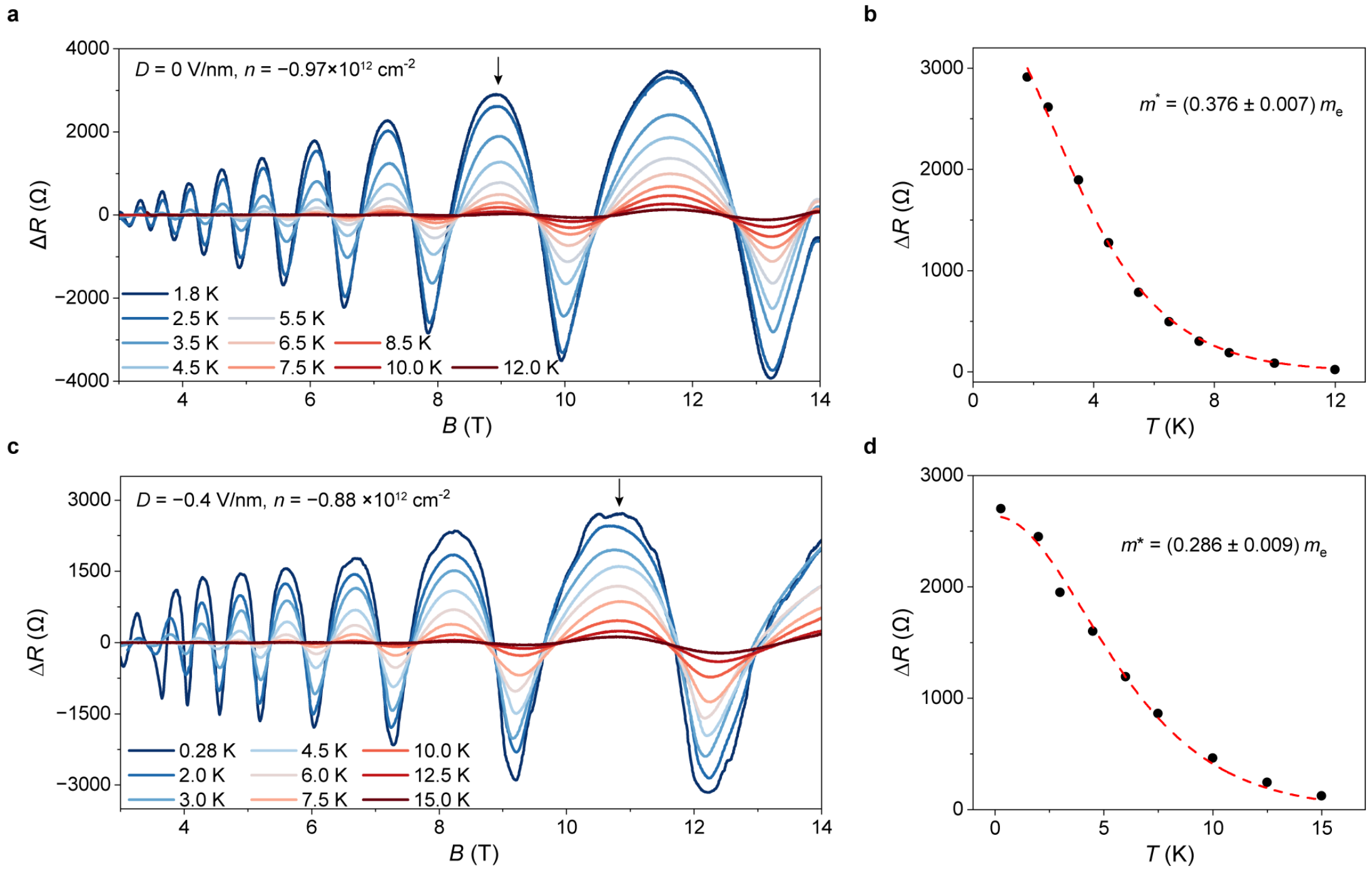

**Extended Data Fig. 3 | Extraction of effective mass via temperature-dependent Shubnikov-de Haas (SdH) oscillations. a,c,** Temperature dependence of the longitudinal resistance oscillations $\Delta R$ as a function of magnetic field $B$ at two different displacement fields $D$ and carrier densities $n$. **a**, $D = 0$ V/nm, $n = -0.97 \times 10^{12}$ cm$^{-2}$, corresponding to region $I_2$; **c,** $D = -0.4$ V/nm, $n = -0.88 \times 10^{12}$ cm$^{-2}$, corresponding to region $I_3$. **b,d,** Temperature-dependent oscillation amplitudes extracted from the specific peaks (indicated by black arrows in **a** and **c**) within the spin-polarized state. The red dashed lines represent fits to the Lifshitz-Kosevich (L-K) formula. The extracted effective masses are $m^* = (0.376 \pm 0.007)\ m_e$ for **b** and $m^* = (0.286 \pm 0.009)\ m_e$ for **d**, where $m_e$ is the free electron mass.

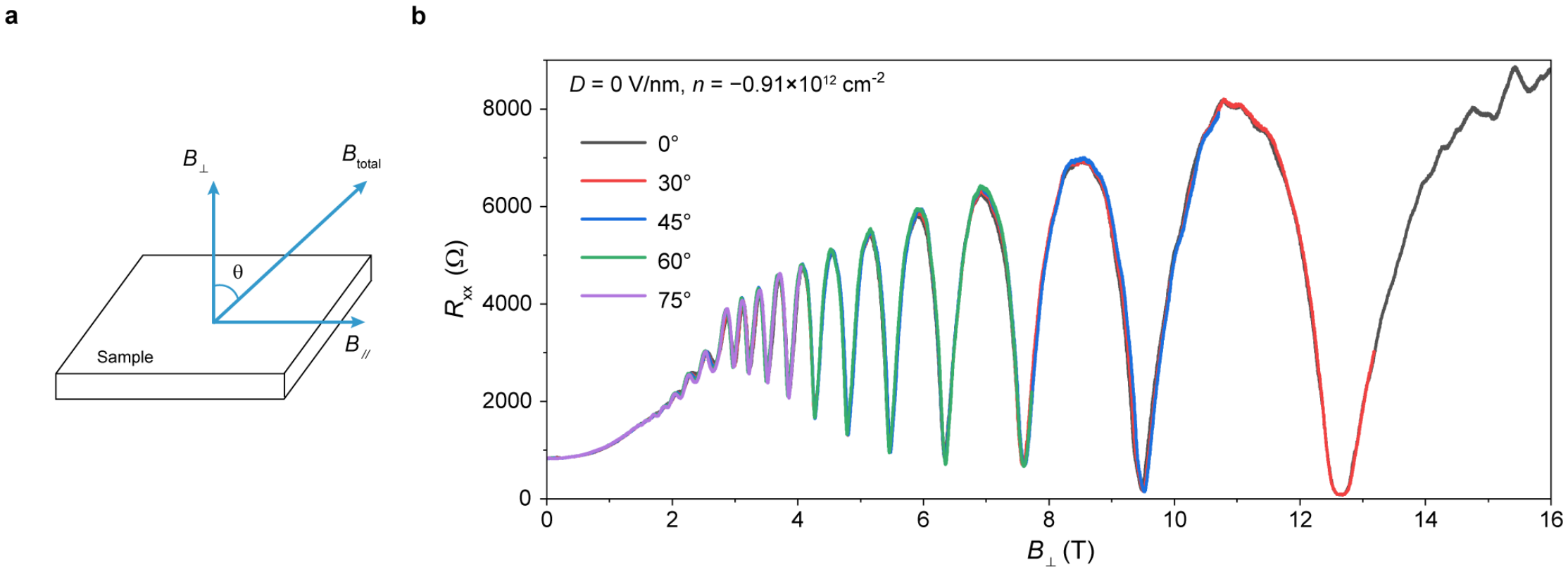

**Extended Data Fig. 4 | Angle-dependent SdH oscillations in HTWSe$_2$. a,** Schematic of the device geometry and magnetic-field configuration. The tilt angle $\theta$ is defined as the angle between the total magnetic field $B_{\text{total}}$ and the sample normal. The field is decomposed into out-of-plane and in-plane components, $B_{\perp}$ and $B_{/\!/}$, respectively. **b,** Longitudinal resistance $R_{xx}$ measured at different tilt angles, plotted as a function of $B_{\perp}$, at $D = 0$ V/nm and $n = -0.91 \times 10^{12}$ cm$^{-2}$ (corresponding to middle-layer-polarized region I$_2$). The SdH oscillations measured at different angles nearly overlap when plotted against the same out-of-plane field component.

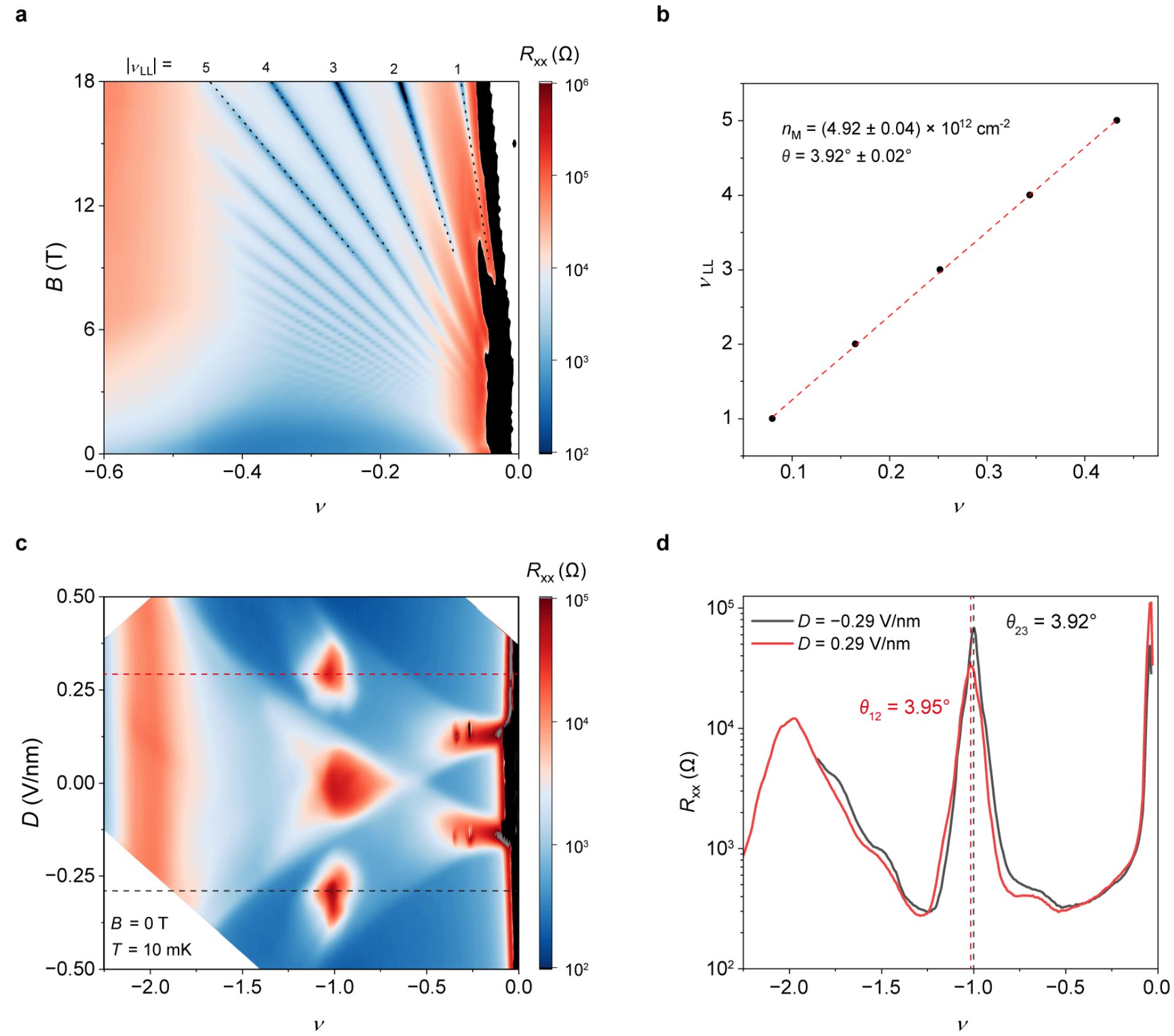

**Extended Data Fig. 5 | Twist-angle calibration. a,** Landau fan diagram of the device at $D = 0$, showing longitudinal resistance $R_{xx}$ as a function of filling factor $\nu$ and magnetic field $B$. Well-defined Landau level sequences emanate from band edge. **b,** Linear dependence of Landau level index $\nu_{LL}$ on filling factor $\nu$, extracted from the $B = 18$ T data in **a**, yielding a carrier density $n_M = (4.92 \pm 0.04) \times 10^{12}$ cm$^{-2}$ and an estimated twist angle $\theta = 3.92 \pm 0.02°$. **c,** Longitudinal resistance $R_{xx}$ as a function of filling factor $\nu$ and displacement field $D$ at $B = 0$ and $T = 10$ mK. Distinct correlated features appear around fractional and integer fillings. **d,** Line cuts of $R_{xx}(\nu)$ taken along the two dashed lines in **c** at $D = -0.29$ V/nm and $D = 0.29$ V/nm. The resistance peak near filling $\nu = -1$ (dashed lines) allows extraction of the effective moiré densities for the 1-2 and 2-3 layer pairs, yielding a small twist-angle difference between the two interfaces ($\theta_{12} \approx 3.95°$, $\theta_{23} \approx 3.92°$).

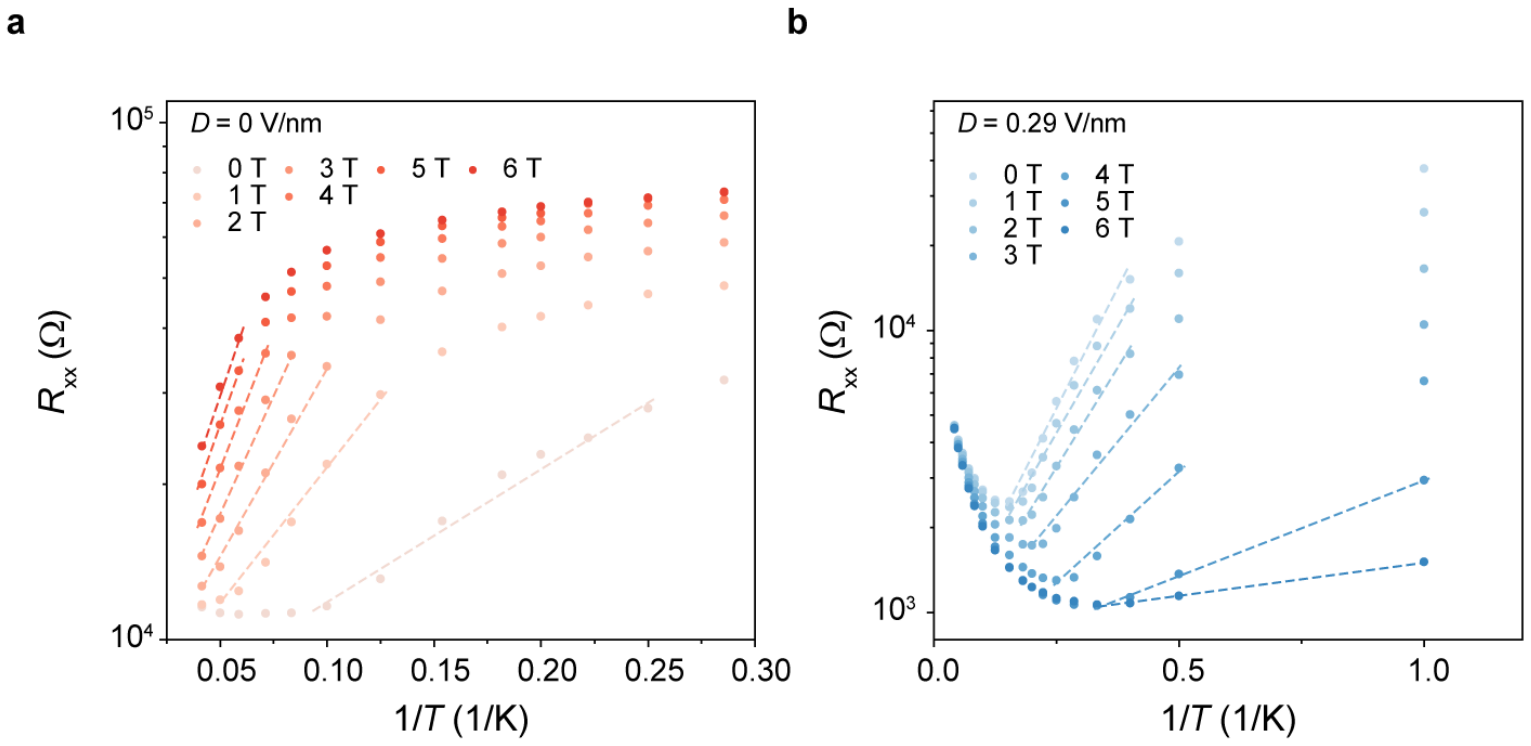

**Extended Data Fig. 6 | Thermal activation analysis. a,** Arrhenius plots of the longitudinal resistance $R_{xx}$ at $D$ = 0 V/nm under selected magnetic fields. Dashed lines are linear fits to the activated transport regime, from which the activation gap $\varepsilon$ is extracted. **b,** Same as in **a**, but for $D$ = 0.29 V/nm. The activated behavior is progressively weakened with increasing magnetic field, consistent with the suppression of the finite-$D$ insulating state. The extracted gap values are summarized in the insets of Fig. 3b and 3e.

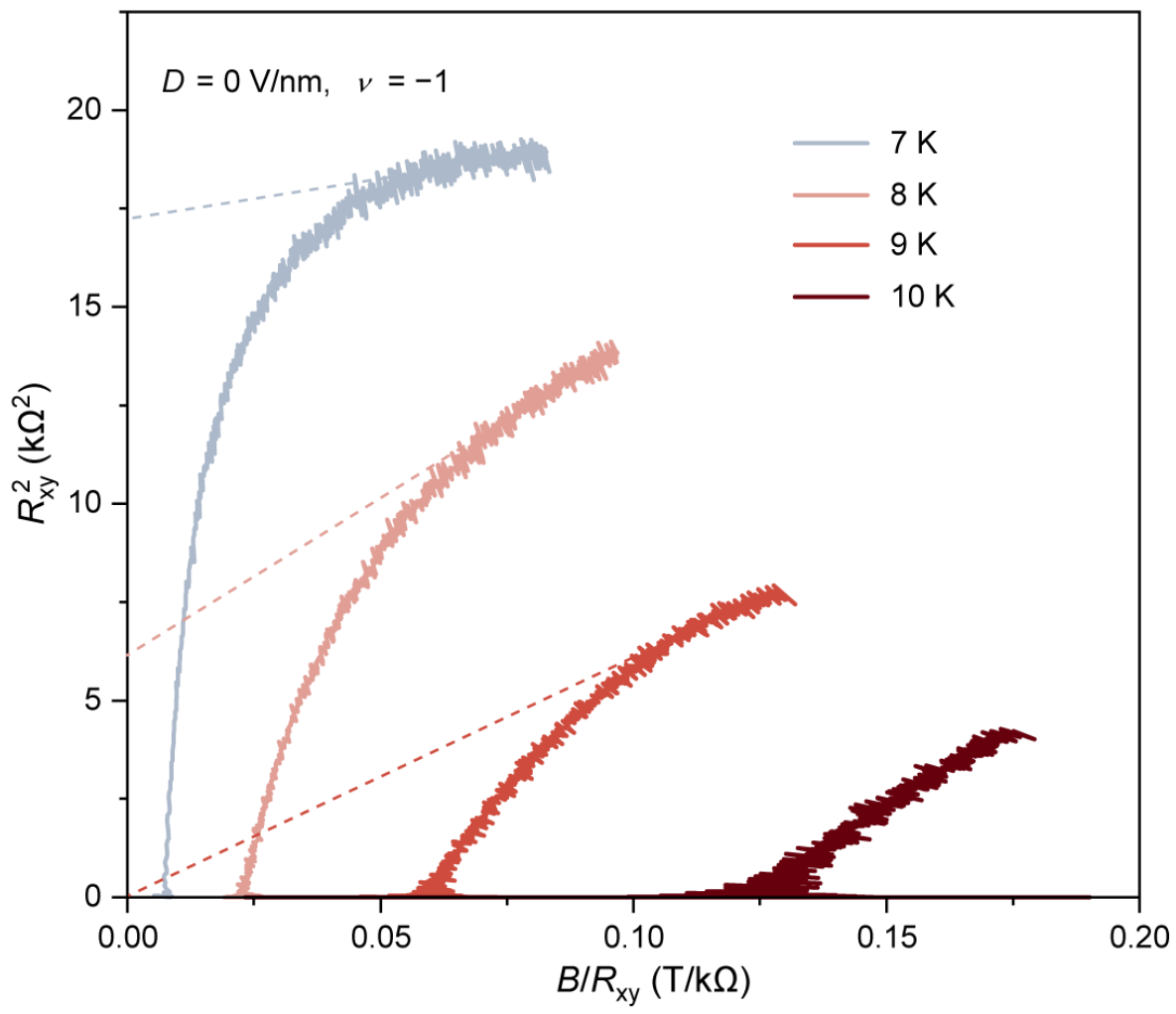

**Extended Data Fig. 7 | Extraction of the Curie temperature.** $R_{xy}^2$ plotted as a function of $|B/R_{xy}|$ at selected temperatures for the $\nu = -1$ state at $D = 0$ V/nm. Linear extrapolation is used to distinguish the magnetic character: a positive intercept indicates a ferromagnetic state, whereas a negative intercept indicates a paramagnetic state. The intercept approaches zero near $T \approx 9$ K, yielding an estimated Curie temperature of about 9 K.

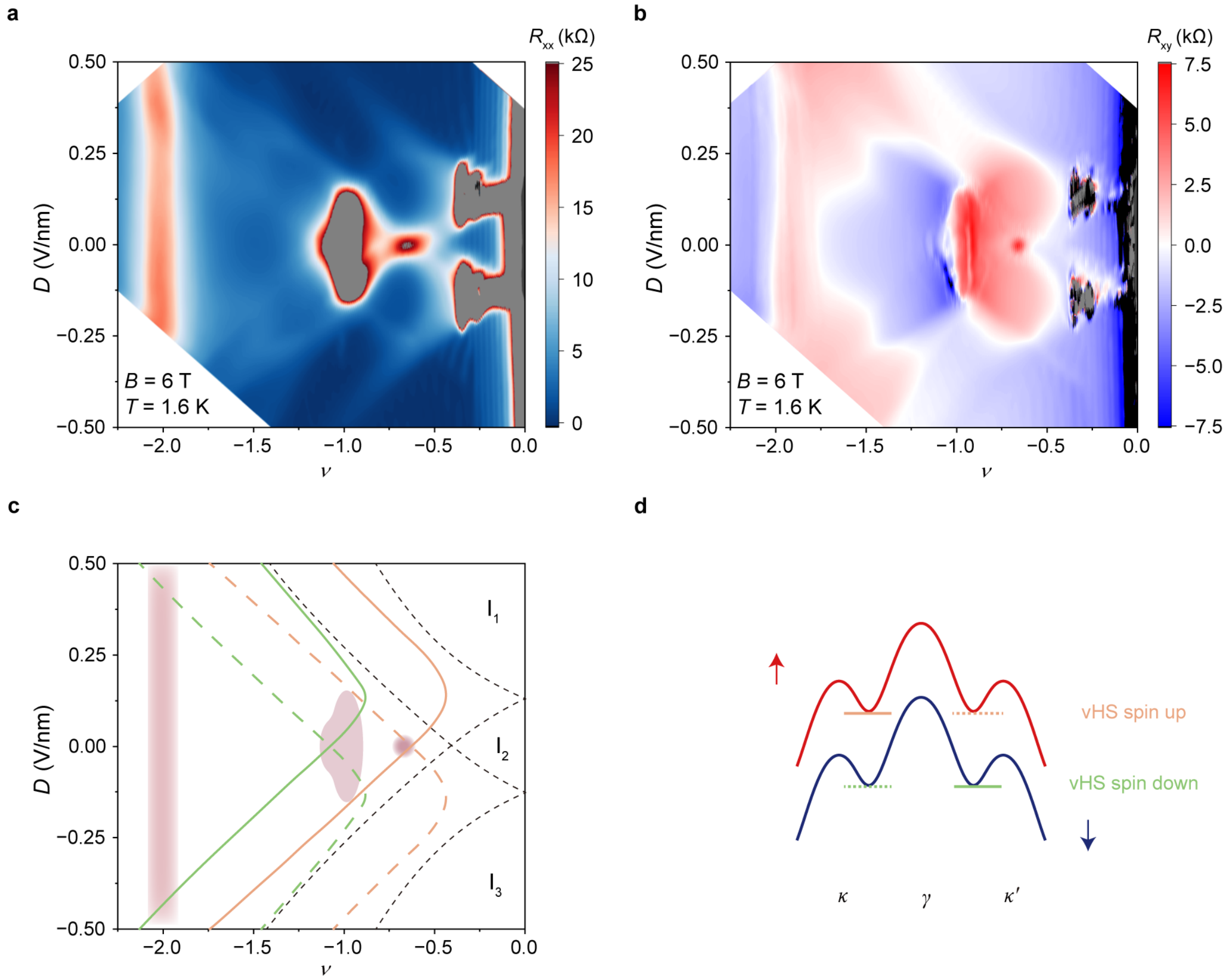

**Extended Data Fig. 8 | Magnetic-field-induced shift of van Hove singularities. a,b,** Longitudinal resistance $R_{xx}$ (**a**) and Hall resistance $R_{xy}$ (**b**) plotted as functions of filling factor $v$ and displacement field $D$ at $B$ = 6 T and $T$ = 1.6 K. **c,** Van Hove singularities extracted from the features in **a** and **b**. Orange and green mark the vHS associated with the spin-up and spin-down branches, respectively, while solid and dashed lines denote vHS derived from the moiré 12 and moiré 23 subsystems. **d,** Schematic of the moiré band structure at $D$ = 0 under magnetic field, illustrating the field-induced splitting between the $K$-valley (spin-up, red) and $K'$-valley (spin-down, blue) bands. The labelled vHS positions correspond to the branches tracked in **c**.

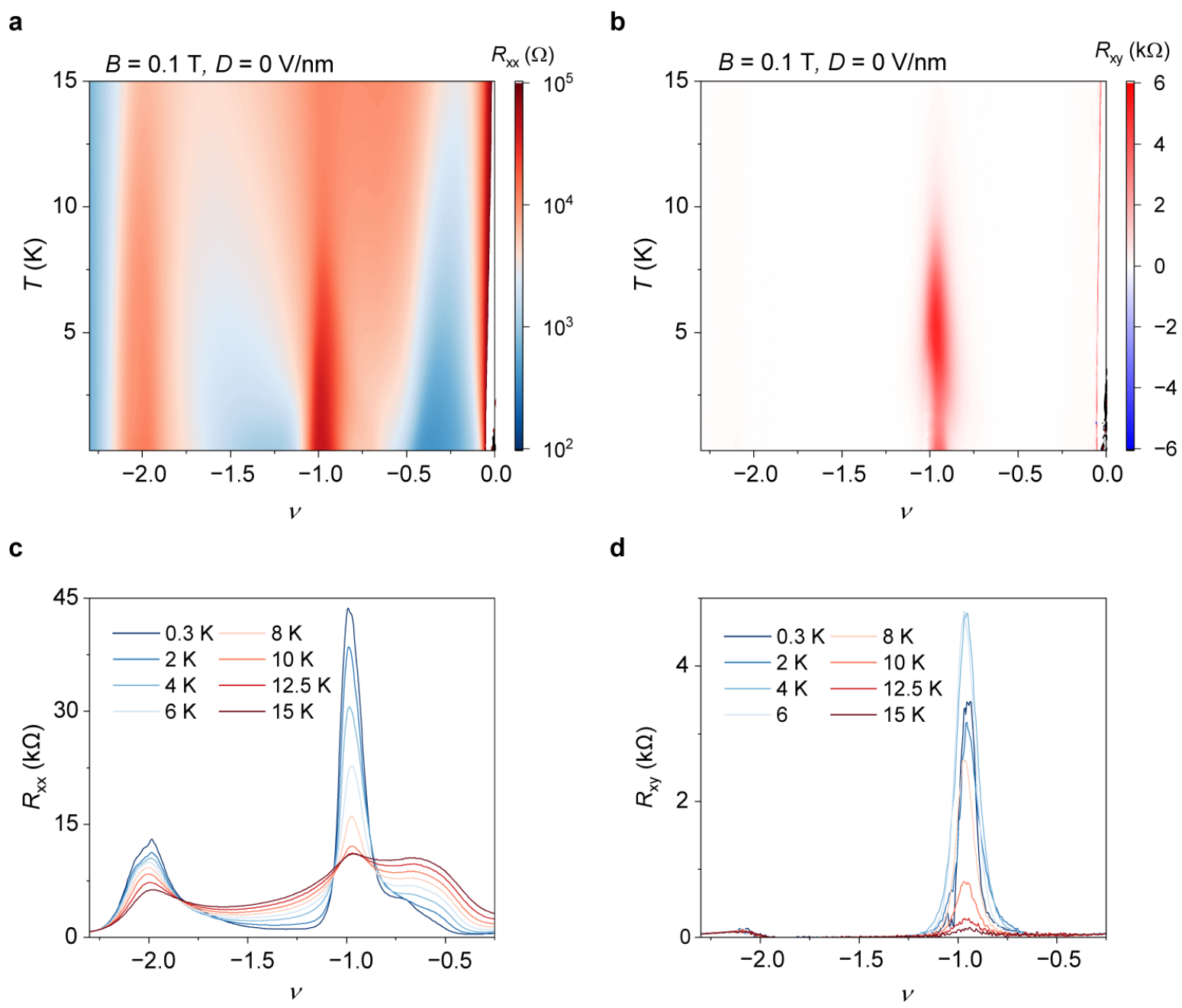

**Extended Data Fig. 9 | Temperature Dependence of Transport Characteristics at Zero Displacement Field. a,b,** Longitudinal resistance $R_{\mathrm{xx}}$ (**a**) and Hall resistance $R_{\mathrm{xy}}$ (**b**) as a function of filling factor $\nu$ and temperature $T$ at $B = 0.1$ T and $D = 0$ V/nm. A prominent resistance peak is observed near $\nu = -1$, corresponding to the ferromagnetic insulating state. **c,d,** Line cuts of $R_{\mathrm{xx}}$ (**c**) and $R_{\mathrm{xy}}$ (**d**) at various temperatures, extracted from **a** and **b**.

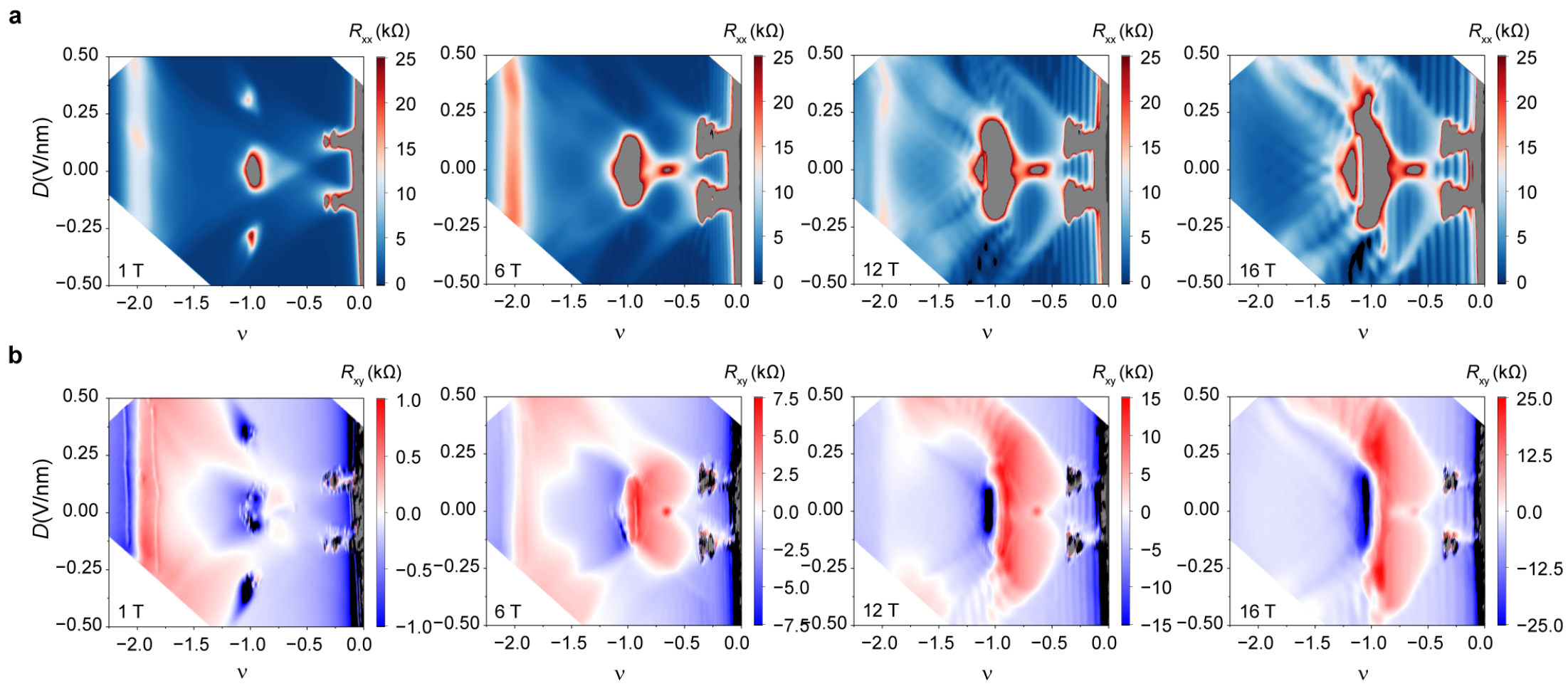

**Extended Data Fig. 10 | Magnetic-field evolution of the *ν*-*D* maps. a,** Longitudinal resistance $R_{xx}$ plotted as a function of filling factor $\nu$ and displacement field $D$, measured at $T$ = 1.6 K and $B$ = 1, 6, 12, and 16 T (from left to right). **b,** Corresponding Hall resistance $R_{xy}$ measured under the same conditions. Each panel uses an individual color scale.